\documentclass[twocolumn,showpacs,preprintnumbers,amsmath,amssymb,superscriptaddress,prd]{revtex4}

\usepackage{graphicx}
\usepackage{epstopdf}
\usepackage{dcolumn}
\usepackage{bm}

\def\naive{na\"{\i}ve}
\def \figabbr{Fig.\ }
\def \eqnabbr{Eq.\ }
\def \secabbr{Section\ }

\def \ecm{$e\,$cm}

\def \vecB{{\bf B}_0}
\def \vecE{{\bf E}}

\def \dBzdz{\partial B_z/\partial z}

\def \vxE{{\bf v}\times{\bf E}}
\def \a16{a_{16}}
\def \chinu{\chi^2/\nu}

\def \tablshielddimensions{I}

\begin{document}

\title{Apparatus for Measurement of the Electric Dipole Moment of the Neutron using a Cohabiting Atomic-Mercury Magnetometer}

\author{C.A. Baker}\affiliation{Rutherford Appleton Laboratory, Chilton, Didcot, Oxon OX11 0QX, UK}
\author{Y. Chibane}\affiliation{University of Sussex, Falmer, Brighton BN1 9QH, UK}
\author{M. Chouder}\affiliation{University of Sussex, Falmer, Brighton BN1 9QH, UK}
\author{P. Geltenbort}\affiliation{Institut Laue-Langevin, BP 156, F-38042 Grenoble Cedex 9, France}
\author{K. Green}\affiliation{Rutherford Appleton Laboratory, Chilton, Didcot, Oxon OX11 0QX, UK}
\author{P.G. Harris}\affiliation{University of Sussex, Falmer, Brighton BN1 9QH, UK}
\author{B.R. Heckel}\affiliation{Department of Physics, University of Washington, Seattle, WA 98195} 
\author{P. Iaydjiev\footnote{On leave of absence from INRNE, Sofia, Bulgaria}}\affiliation{Rutherford Appleton Laboratory, Chilton, Didcot, Oxon OX11 0QX, UK}
\author{S.N. Ivanov\footnote{On leave of absence from Petersburg Nuclear Physics Institute, Russia}\footnote{Present address: Institut Laue-Langevin, BP 156, F-38042 Grenoble Cedex 9, France}}\affiliation{Rutherford Appleton Laboratory, Chilton, Didcot, Oxon OX11 0QX, UK}
\author{I. Kilvington}\affiliation{Rutherford Appleton Laboratory, Chilton, Didcot, Oxon OX11 0QX, UK}
\author{S.K. Lamoreaux\footnote{Present address: Department of Physics, Yale University, New Haven, CT 06520}}\affiliation{Department of Physics, University of Washington, Seattle, WA 98195} 
\author{D.J. May}\affiliation{University of Sussex, Falmer, Brighton BN1 9QH, UK}
\author{J.M. Pendlebury}\affiliation{University of Sussex, Falmer, Brighton BN1 9QH, UK}
\author{J.D. Richardson}\affiliation{University of Sussex, Falmer, Brighton BN1 9QH, UK}
\author{D.B. Shiers}\affiliation{University of Sussex, Falmer, Brighton BN1 9QH, UK}
\author{K.F. Smith\footnote{deceased}}\affiliation{University of Sussex, Falmer, Brighton BN1 9QH, UK}
\author{M. \surname{van der Grinten}}\affiliation{Rutherford Appleton Laboratory, Chilton, Didcot, Oxon OX11 0QX, UK}

\date{\today}

\begin{abstract}
A description is presented of apparatus used to carry out an experimental search for an
electric dipole moment of the neutron, at the Institut Laue-Langevin (ILL), Grenoble. The experiment incorporated a cohabiting atomic-mercury magnetometer in order to reduce spurious signals from magnetic field fluctuations.  The 
result has been published in an earlier letter \cite{baker06}; here, the methods and
equipment used are discussed in detail. \end{abstract}

\maketitle
\section{Introduction}

\subsection{The electric dipole moment of the neutron}

Any non-degenerate system of defined, non-zero angular momentum will have a
permanent electric dipole moment (EDM) $d$ if its interactions are
asymmetric under both parity (P) and time (T) inversion \cite{ramsey82,barr89,sachs_time_rev}. The neutron carries
spin $\frac12$, and it also possesses the virtue of being sensitive to all
known particle physics interactions. It is therefore expected to possess a
finite EDM with its magnitude dependent upon the nature and origin of the T
violation, and this EDM is, in turn, a sensitive probe of such asymmetric
interactions.

Parity violation \cite{wu57} is a well-established property of the weak interaction in
general.  Evidence for $T$ violation, which arises at a much weaker level, has come from the observation that
there is a (0.66 $\pm $ 0.18) \% greater probability for a $\overline{K}^{0}$
to turn into a $K^{0}$ than the other way around \cite{angelopoulos98}, and that there is an angular asymmetry in the rare decay $%
K_{L}\rightarrow \pi ^{+}\pi ^{-}e^{+}e^{-}$ of $\left( 14.6\pm 2.3\pm
1.1\right)$\% \cite{alavi99a,adams98}.    T violation
and CP violation, where C is charge conjugation, are closely related through
the CPT theorem \cite{luders54,schwinger53a,schwinger53b} which predicts the invariance of the combined symmetry. Any CP violation in a CPT-invariant theory therefore implies the breakdown of
time-reversal symmetry and leaves a finite expectation value of the neutron
EDM. Violation of CP-symmetry has been studied in detail in the $K^{0}$
system \cite{christenson64} and, more recently, in the $B$ system \cite{babar01, abe01}; see, for example, \cite{cpreview04} 
and references therein. 

The origins of CP violation are still unknown. In the kaon system it is dominated by indirect ($\Delta S=2$) contributions due to mixing. It has been observed  
\cite{alavi99b,barr93a} in direct
quark interactions ($\Delta S=1$).  Contributions from ``superweak'' $\Delta S=2$ interactions specific to the kaon systems have been ruled out.  

\begin{figure} [ht]
\begin{center}
\resizebox{0.5\textwidth}{!}{\includegraphics{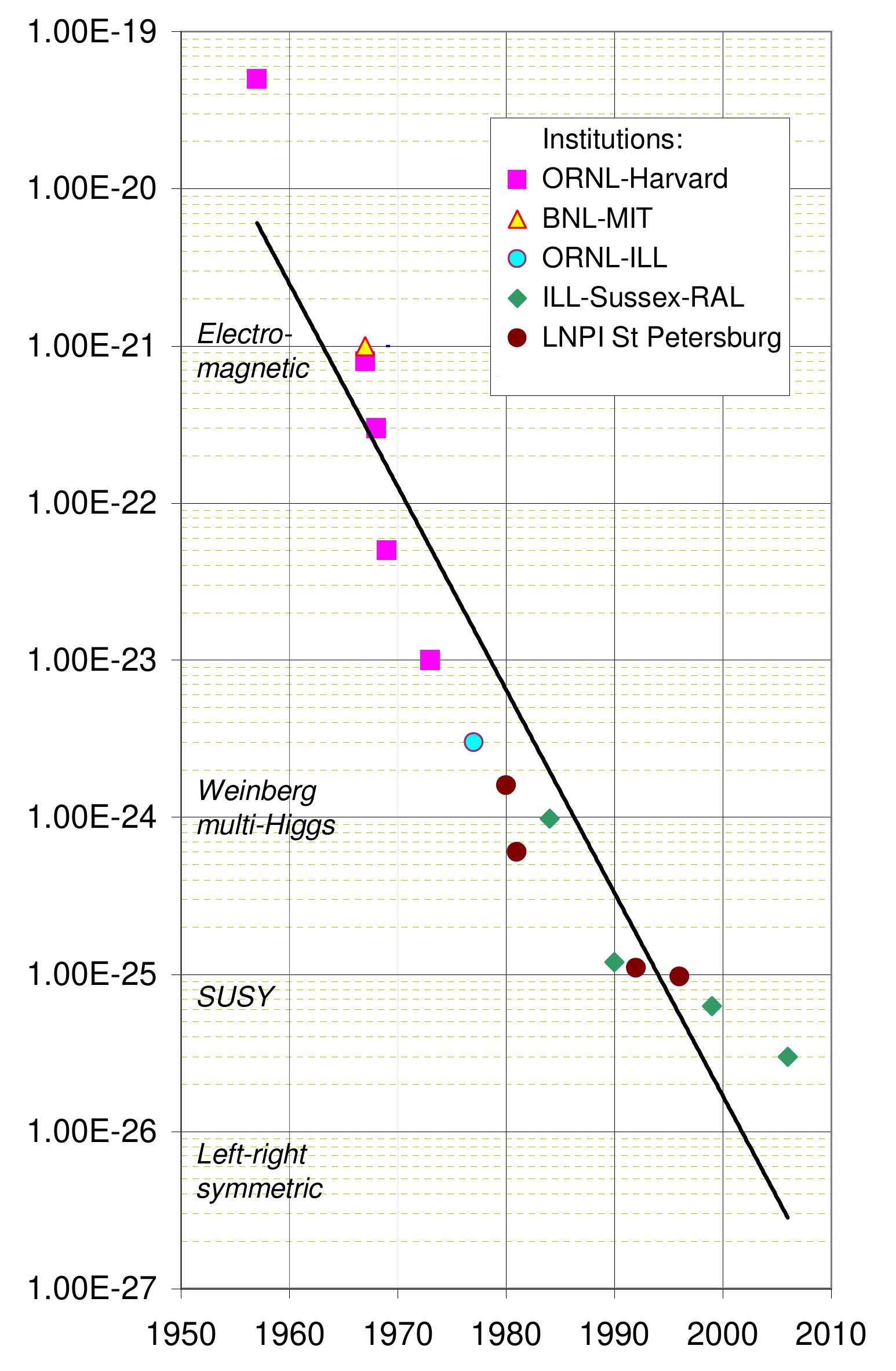}}
\end{center}
\caption{(Color online) The evolution of the experimental limit of the
electric dipole moment of the neutron. Those experiments before 1980 used
neutron beams, and those after use stored ultracold neutrons. See \cite{ellis89} for the theoretical predictions.}
\label{edm_limits}
\end{figure}

Many alternative theories exist (see, for example,
contributions in \cite{jarlskog89}), but the data from the $K^{0}$ and $b$ systems
alone are insufficient to distinguish between them. These theories also
predict non-zero values for the EDM of the neutron, but the predictions
differ, one from another, by many orders of magnitude \cite{ellis89}. The major difference between the theories is that in some, and in
particular in the standard SU(2) $\times $ U(1) model of electroweak
interactions, the contributions to the EDM appear only in second order in
the weak interaction coupling coefficient, whereas in others the
contributions are of first order in the weak interaction. Detection of the
latter larger size of EDM would be evidence for new physics beyond the
standard model \cite{he89}. 
The small size of the neutron EDM, as
indicated by the measured values displayed in \figabbr~\ref{edm_limits} 
\cite{smith57,miller67,shull67,dress68,%
cohen69,baird69,apostolescu70,dress73,%
dress77,altarev80,altarev81,pendlebury84,%
altarev86,smith90,altarev92,altarev96,harris99,baker06}, has already eliminated many
theories, and is pressing heavily upon the expectations from extensions to
the Standard Model through to supersymmetric interactions. 

\subsection{Implications of non-zero EDM measurements}

EDMs are being sought in various systems: the free neutron, the mercury atom \cite{romalis01}, and the electron (via the thallium atom \cite{regan02} and, more recently, the YbF \cite{hudson02} and PbO molecules \cite{kozlov02}), in addition to a proposal to study deuterium \cite{lebedev04}.  The fundamental mechanisms underlying sources of EDMs are different in each system, and the measurement of a finite value within one of these systems would therefore have distinctive implications \cite{abel06}: For example, if the EDMs are driven by the QCD  $\theta$ angle, one 
would expect similar contributions to all strongly coupled systems, with 
the consequent approximate pattern $d_n\sim d_{Hg}\sim d_D >> d_{e},$ and so on.  Thus, the different systems have different implications for physics models beyond the standard model. Measurements on multiple systems are also needed in order to rule out cancellations.  

EDM limits provide fairly tight constraints upon supersymmetric models; the same is true of most other models beyond the standard model that attempt to incorporate $CP$ violation to a degree adequate to explain the observed baryon asymmetry of the Universe.  The ``accidental'' cancellation of first-order contributions in the Standard Model is not a general feature, and EDM limits (and EDM values, once measured) provide a powerful way to distinguish between models and, indeed, to eliminate many of them.  Ramsey \cite{ramsey95} and Barr \cite{barr93a} have provided useful reviews of the situation, and the book by Khriplovich and Lamoreaux \cite{khriplovich_lamoreaux} contains further general information on EDMs.

\section{Principle of the method}
\label{sec:principle}

Almost all of the experimental searches for the EDM of the neutron
have been magnetic resonance experiments in which polarized neutrons are
subjected to parallel magnetic and electric fields in vacuum \cite{shapiro68},\cite{golub72}. The only internal degrees of freedom of
the neutron are those associated with the spin ${\bf s}$, so that the
Hamiltonian ($\mathcal{H}$) in an electric (${\bf E}_{0}$) and a magnetic (${\bf B}_{0}$) field is 
\begin{equation}
\mathcal{H}=-2{\bf s}\cdot (\mu _{n}{\bf B}_{0}+d_{n}{\bf E}_{0}).
\end{equation}
If the magnetic and electric fields are parallel or antiparallel, the
precession frequency $\nu _{0}$ of the spin is given by 
\begin{equation}
\label{eqn:hnu}
h\nu _{0}=-2\mu _{n}|{\bf B}_{0}|\mp 2d_{n}|{\bf E}_{0}|,
\end{equation}
where $h$ is Planck's constant, $\mu _{n}$ is the magnetic dipole moment ($
-1.913\ldots $\ nuclear magnetons), $d_{n}$ is the EDM
and the upper (lower) sign is for $\vecB$ and $\vecE$ parallel
(antiparallel). When an
electric field of magnitude $E_{0}$ is changed from being parallel to $\vecB$
to being antiparallel, the precession frequency changes by 
\begin{equation}
\delta \nu _{0}=-{\frac{4d_{n}E_{0}}{h}}.
\label{eqn: freq shift prop to E field x edm}
\end{equation}
An EDM of $10^{-25}$ \ecm\  would give a frequency shift of 1~$\mu $
Hz with the reversal of a 1 MV/m electric field. Because $\mu_n$ is negative, the sign
definition for $d_{n}$ is such that a positive dipole moment would increase
the precession frequency when $\vecE$ and $\vecB$ are antiparallel. 
Application of a magnetic field produces a magnetic Zeeman
splitting; subsequent application of an electric field then merely changes
the separation of the Zeeman levels, without inducing any further splitting.
It should be noted that the electric polarizability of the neutron cannot
affect the precession frequency to first order.

The early experiments used beams of neutrons with velocities greater than 100 m/s. Such experiments became limited by the $\vxE$ effect, according to which motion through the electric field
results in a magnetic field in the neutron rest frame and hence a possible
change in the precession frequency with the same dependence on the electric
field as a real EDM. More recent experiments use ultra-cold neutrons (UCN),
with velocities of less than 7~m/s, stored in evacuated chambers with
walls that totally reflect the neutrons; the average velocity is so close to
zero that the $\vxE$ effect can be adequately controlled at
the present level of sensitivity. The first published result from a series
of experiments being carried out under these conditions at the Institut
Laue-Langevin (ILL) in Grenoble was $d_{n}=-(3\pm 5)\times 10^{-26}$ \ecm\  \cite{smith90}. A broadly similar experiment at the PNPI in Russia \cite{altarev96} yielded an EDM of $(+2.6\pm 4.0\pm 1.6)\times
10^{-26}$ $e$ cm. Both experiments were limited at the time by systematic
uncertainties associated with instabilities and non-uniformities in the
magnetic field. The ILL experiment initially used three rubidium
magnetometers adjacent to the storage cell to try to compensate for
magnetic field drifts; the PNPI experiment used instead a back-to-back
twin-cell arrangement to make simultaneous measurements with the $\vecE$
field in opposite directions. In each case, the presence of gradients in the
magnetic field could adversely affect the results, since there was a
significant displacement between each measurement cell and the control
volume used for compensation. This problem was addressed in this experiment at the ILL
by the installation of a magnetometer based upon measurement of the precession frequency of spin-polarized $I=1/2$ atoms of $^{199}$Hg ($3\times
10^{10}$ atoms/cm$^{3}$; $\mu_n/\mu_{Hg} = \gamma_n/\gamma_{Hg} = -3.842$) stored simultaneously in the same trap as
the neutrons.  Using \eqnabbr (\ref{eqn:hnu}) for both the neutrons and the mercury, and assuming that both experience the same $B$, we find that to first order in the EDMs $d$, 
\begin{equation}
\label{eqn:freq_ratio}
\frac{{\nu_n }}{{\nu _{Hg} }} = \left| {\frac{{\gamma _n }}{{\gamma _{Hg} }}} \right| + \frac{{(d_n  + \left| {\gamma _n/\gamma _{Hg} } \right|d_{Hg} )}}{{\nu _{Hg} }}E = \left| {\frac{{\gamma _n }}{{\gamma _{Hg} }}} \right| + \frac{{d_{meas} }}{{\nu _{Hg} }}E.
\end{equation}

It is worth noting that \eqnabbr (\ref{eqn:freq_ratio}) is only valid in a non-rotating reference frame.  The rotation of the Earth imparts a small but perceptible shift in this frequency ratio \cite{lamoreaux07}.

For each data-taking run, the measured EDM $d_{meas}$ was obtained from a linear fit to the ratio $\nu_n / \nu_{Hg}$ versus $E$.  \eqnabbr (\ref{eqn:freq_ratio}) shows that $d_{meas}$ contains a contribution from $d_{Hg}$. The true $d_{Hg}$ has been shown to be $(-1.06\pm0.49\pm0.40)\times 10^{-28}$ \ecm\  \cite{romalis01}, so it introduces a systematic error of $(-0.4\pm0.3)\times10^{-27}$ \ecm\ into $d_{meas}$.

     To the true $d_n$ and $d_{Hg}$ within $d_{meas}$ there will also be added coefficients of fractional shifts in $\nu_n$ and $\nu_{Hg}$, from other causes, which are linear in $E$ and thus constitute additional systematic errors.  The most important of these involves a geometric phase (GP) arising when the trapped particles experience a gradient $\partial B_{0_z}/\partial z$ in the presence of $E$ \cite{pendlebury04}. This particular effect has now been characterised and understood, and to a large extent it has been possible to compensate for it.


\section{Ultracold neutrons}

As a consequence of the coherent strong interaction between neutrons and the
nuclei of a material medium, the surface of the medium presents a potential
step relative to vacuum for long-wavelength neutrons. This potential $V_{F}$, 
called the mean Fermi potential, is given by \cite{golub79} 
\begin{equation}
V_{F}={\frac{2\pi \hbar ^{2}}{m}}Nb,
\end{equation}
where $m$ is the mass of the neutron and $N$ the number of atoms per unit
volume with mean coherent forward scattering length $b$. A neutron with velocity
less than the critical velocity $v_{c}$, defined by $mv_{c}^{2}/2=V_{F}$,
will be reflected from the surface for any angle of incidence. The Fermi
potential for most materials is less than 300~neV, which corresponds to
critical velocities of less than 7.6 m/s. Such slow
neutrons can be confined in material traps by total external reflection,
and are called ultra-cold neutrons (UCN). Nuclear reactors are a source of
UCN, which constitute the very low energy part of the spectrum of moderated
neutrons.

For cold and ultra-cold neutrons in a magnetic material the Fermi potential
due to the nuclear scattering acquires an additional term representing the
interaction of the magnetic moment of the neutron $\mu _{n}$ with the
internal magnetic field B of the material. Thus, 
\begin{equation}
V_{F}={\frac{2\pi \hbar ^{2}}{m}}Nb\pm \mu _{n}B,
\end{equation}
where the $\pm $ refers to the two spin states of the neutron. It is
possible to find ferromagnetic materials with very low Fermi potentials for
one spin state of the neutron and high Fermi potentials for the other spin
state. It is then possible to spin-polarize UCN by transmission through a
thin magnetised foil of such a material.

In the experiment described in this paper the number density of UCN was less
than 10 cm$^{-3},$ and hence neutron-neutron collisions were
extremely unlikely and can be ignored.

\subsection{Upscattering and absorption of UCN in materials}

Although the UCN have speeds characteristic of a temperature of about 2~mK,
the neutron storage trap was maintained at room temperature. At first
sight it might appear surprising that these neutrons could be stored for
hundreds of seconds without being scattered out of the UCN energy range.
This was possible because the thermal motions of individual nuclei in the
walls of the trap were sensed only weakly by the UCN, which were reflected
by the combined coherent scattering from millions of nuclei lying within a
short distance (of the order of 100 \AA ) of the surface. In this coherent
scattering, the thermal motion of the center of mass of such a large group
was negligible compared with the speed of the UCN. At the same time, any
recoil energy associated with the group was also negligible. In addition, 
collisions that involved an exchange of energy with a smaller group of
nuclei in the wall, and hence an upscattering of the neutron out of the UCN
energy range, were infrequent (although important in determining the mean storage lifetime): the following argument has been
given by Zeldovich \cite{zeldovich59}. When a neutron is reflected from a
surface, its wave function penetrates into the wall a distance of order $%
(\lambda /2\pi )$, where $\lambda $ is the de~Broglie wavelength. In a
storage volume of dimension $l$, each neutron with velocity $v$ which is
stored for a time $T_s$ accumulates a total path length $L$ inside the
material of the walls, where 
\begin{equation}
L\approx {\frac{\lambda T_sv}{2\pi l}}.
\end{equation}
For the typical values of $T_s=150$~s, speeds $v$ of up to about 5 m/s, and $l=150$%
~mm, a value of $L=60~\mu \mathrm{m}$ is obtained. This distance is
sufficiently small, compared with observed UCN interaction lengths, that one
expects very little inelastic scattering and absorption of the neutrons.

In general, the survival times of UCN in material traps, particularly
those made from materials with low absorption cross sections, are less than
would be calculated for pure materials. This is caused by the presence of
impurities (particularly hydrogen) in the surface, which drastically reduces
the survival time \cite{stoika78,lanford77}. To reduce
hydrocarbon contamination of the trap used in this EDM experiment, the majority
of pumps in the vacuum system were oil-free turbopumps; the remaining diffusion
pumps were filled with Fomblin \cite{montedison_grp} oil, which is a fully fluorinated
polyether \cite{bates82}. The chemical formula of Fomblin is CF$_{3}$(OCF$_{3}$%
CFCF$_{2}$)$_{m}$(OCF$_{2}$)$_{n}$OCF$_{3}$. To reduce the presence of surface hydrogen still further, the trap surfaces were discharge-cleaned using 1 torr of oxygen.

\subsection{Depolarization in wall collisions}

If neutrons are stored in a trap made of a material with non-zero magnetic
moments, the interaction between a neutron and the wall will be spin
dependent. Collisions with the walls will therefore result in depolarization
of the neutrons. The magnitude of this depolarization can be estimated using
a simple random walk model, similar to that of Goldenberg, Kleppner and
Ramsey \cite{goldenberg61}. If, during one collision with the
wall, the two spin states of the neutron experience Fermi potentials that
differ by $\Delta V_{F}$, and the interaction lasts a time $\tau $, the spin
of a neutron will be rotated through an angle 
\begin{equation}
\delta \phi \approx {\frac{\tau \Delta V_{F}}{\hbar }}.
\end{equation}
For the case where the neutron penetrates a distance $\lambda /2\pi $ into
the wall, 
\begin{equation}
\tau ={\frac{\lambda }{2\pi v}}\approx 2\times 10^{-9}~\mathrm{s}.
\end{equation}
During the storage time the neutron makes $M=T_{s}v/l$ collisions with the
walls, for which the phase shifts, which differ randomly from one wall
collision to another, will add as in a random walk, so that the overall rms
phase shift is 
\begin{equation}
\Delta \phi \approx \overline{\delta \phi }\sqrt{M}={\frac{\overline{\Delta
V_{F}}\lambda }{h}}\sqrt{\frac{T_{s}}{lv}}.
\end{equation}

If the difference in Fermi potentials is $\Delta V_{F}$ for a material in
which all of the nuclei are aligned, a neutron that interacts with $N$
randomly oriented nuclei will experience an average potential difference of $%
\overline{\Delta V_{F}}=\Delta V_{F}/\sqrt{N}$. The number of interacting
nuclei is $N\approx n\left( \lambda /2\pi \right) ^{3},$ and, taking $\Delta
V_{F}=V_{F}=250$~neV, $n=10^{29}~\mathrm{m^{-3}}$, $T_{s}=130$~s, $v=6~%
\mathrm{m\,s^{-1}}$ and $l=150$~mm, a phase difference of 
\begin{equation}
\Delta \phi \approx 0.02~\mathrm{rad}
\end{equation}
is obtained. This implies that polarized neutrons can retain their
polarization for times of the order of $10^{5}~$ s. In practice, the
depolarization time in the storage trap used for this EDM\
experiment was of the order of 600~s.

It follows from the above that the ability to store polarized neutrons is exclusive to storage traps that
are made of non-magnetic materials. If the walls of the trap contain
magnetic domains of size comparable to or greater than the neutron
wavelength, then the interaction of the magnetic moment of the neutrons with
the magnetic field inside the domains dominates. Since the magnetic
interaction can be a few hundred neV, the same size as $V_{F}$, one
effectively suppresses the factor of $1/\sqrt{N}$ in the above calculation
and the neutron polarization survival time drops to values of the order of 50 ms.

\section{Ramsey's method of separated oscillating fields}
\label{sec:ramsey_method}

\begin{figure} [ht]
\begin{center}
\resizebox{0.5\textwidth}{!}{\includegraphics*[clip=true, viewport = 20 129 500 750]{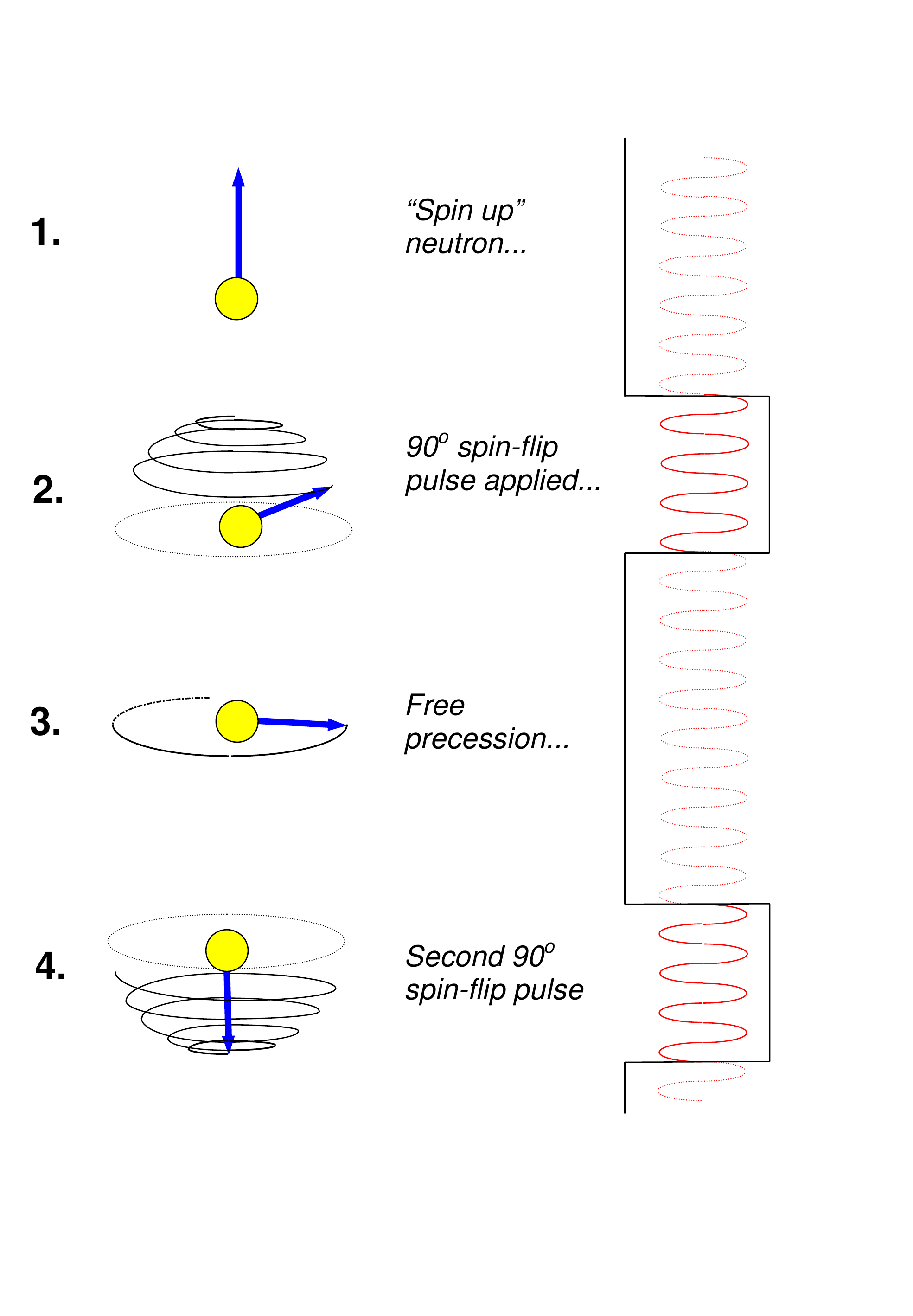}}
\end{center}
\caption{(Color online) The Ramsey method of separated oscillatory fields.  See text for description.}
\label{fig:ramsey_flip}
\end{figure}

The precession frequency of the stored neutrons was determined by the method of separated oscillating fields. The method was devised for
molecular beam experiments where an oscillating field is applied to the beam
at the beginning and at the end of a flight path through an interaction region 
\cite{ramsey_molec_beams, ramsey90}. In this EDM
experiment, where the neutrons were stored in a trap, two short intervals
of phase-coherent oscillating field were applied, one at the beginning and
the other at the end of a period of free precession, so that they were separated in
time but not in space. The phase coherence between the two pulses is achieved by gating off the output
of a single oscillator during the intervening period.  The sequence is shown schematically in \figabbr \ref{fig:ramsey_flip}.

At the start of each measurement cycle within a data-taking run, the neutrons passed through the
magnetised polarizing foil and entered the storage volume with their spin
polarization antiparallel to the uniform magnetic field $\vec{B}_0$ (a state
referred to henceforth as ``spin up''). A resonant oscillating field ${\bf B}_1$, 
perpendicular to $\vecB$ and with a frequency close to resonance,
was applied for 2 seconds with an amplitude such that the neutron
polarization vector was rotated through an angle of $\pi /2$ and brought
perpendicular to $\vecB$. The polarization vector was then left to
precess about $\vecB$ during a period $T_{fp}$ (the subscript here indicating ``free precession''), until the second
phase-coherent oscillating field pulse was applied. If the oscillating field
frequency had been exactly on the center of the resonance, this second pulse 
would have rotated the polarization through a further $\pi /2$ such that it became
parallel to $\vecB$ (the $\hat{z}$ direction), as shown in \figabbr \ref{fig:ramsey_flip}. For frequencies a little
off resonance, the final $\hat{z}$-component of the polarization depends
strongly on the accumulated phase difference between the neutron
polarization vector and the oscillator. When the neutrons were finally
released from storage, the magnetised polarizing foil served as an analyzer,
giving a neutron count that depended linearly upon this final $\hat{z}$
 component of the polarization. Thus, the neutron count was sensitive to the
accumulated precession phase.

Emptying the trap and counting the stored neutrons took 40-50~s. For half
of this time, a 20 kHz oscillating current was applied to a solenoid wrapped
around the guide tube above the polarizer. This flipped the spins of the
neutrons, and allowed the neutrons in the opposite spin state (``spin down'')
to be counted.

\begin{figure} [ht]
\begin{center}
\resizebox{0.5\textwidth}{!}{
{\includegraphics{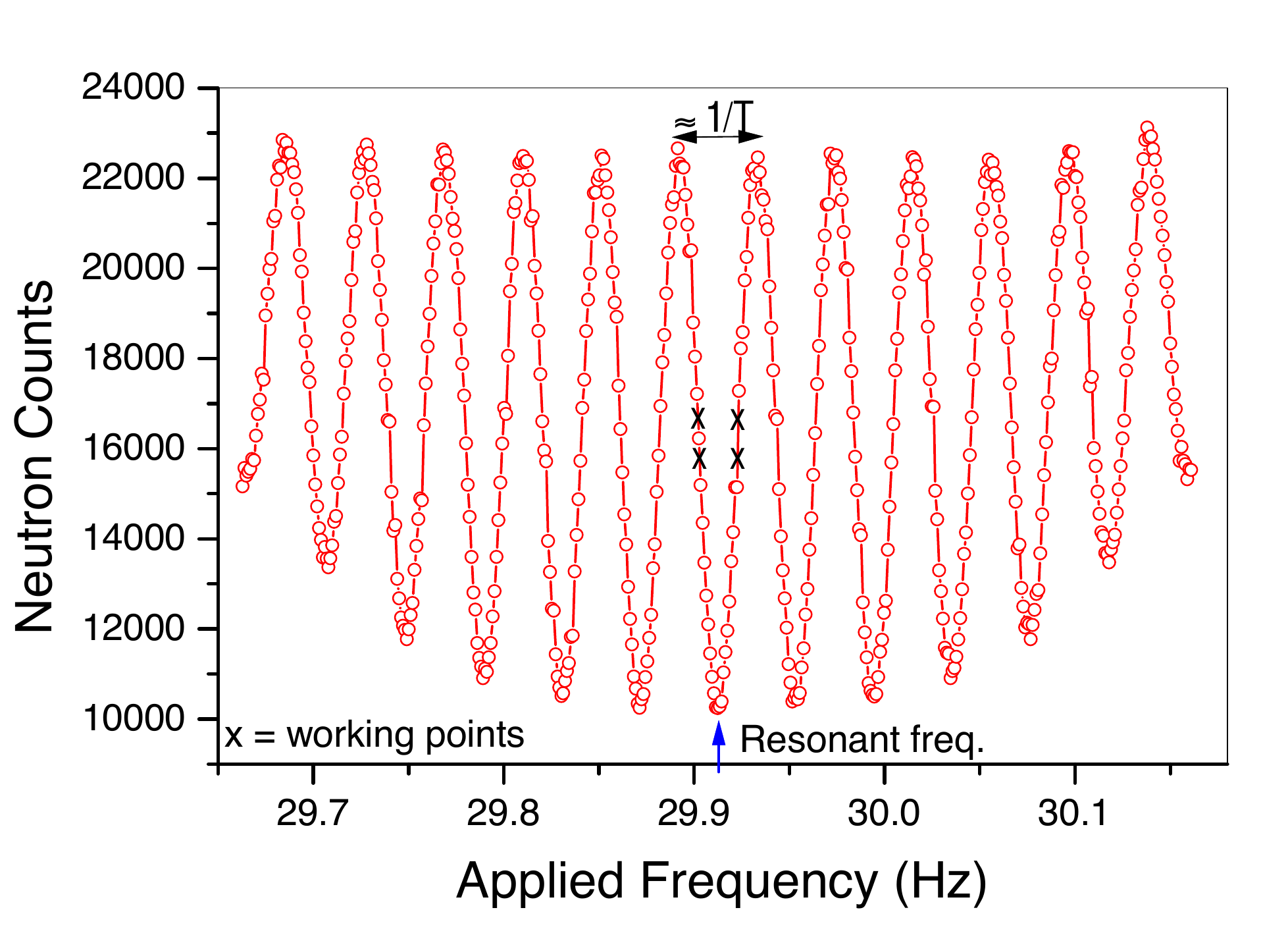}}}

\end{center}
\caption{(Color online)  The Ramsey resonance pattern obtained by
scanning the frequency of the oscillating field ${\bf B}_1$ through the
resonance. The coherence time (between the Ramsey pulses) was 22~s in a 1~$\mu $T magnetic field. 
The ordinate is the number of neutrons in the original spin state counted at
the end of each storage time. Error bars are omitted for clarity. During
normal data taking measurements were taken sequentially at the four points
shown.}
\label{fig:ramsey_resonance}
\end{figure}

\figabbr~\ref{fig:ramsey_resonance} shows the Ramsey resonance pattern
obtained experimentally as the frequency of the oscillating field ${\bf B}_1$ was
varied. It is expected theoretically \cite{ramsey_molec_beams, ramsey90} that, across the central fringes, the number of neutrons counted
as a function of the oscillating field frequency $\nu $ can be described by 
\begin{equation}
N_{\uparrow \downarrow }(\nu )=\overline{N}_{\uparrow \downarrow }\mp \alpha
_{\uparrow \downarrow }\overline{N}_{\uparrow \downarrow }\cos \left( {\frac{%
\pi (\nu -\nu _0)}{\Delta \nu }}\right) ,  \label{eqn: ramsey lineshape}
\end{equation}
where $\overline{N}$ is the average number of neutrons counted for the spin
state in question, up $\uparrow $ or down $\downarrow $. The visibility $
\alpha $ is the product of the neutron polarization and analyzing power,
again for the spin state in question; $\nu _0$ is the resonant frequency,
and the linewidth $\Delta \nu $ is the width at half height of the central fringe. The two
signs $\mp $ also refer to the two spin states. $\overline{N}$ and $\alpha $
(for either spin state) are related to the fringe maximum and minimum $%
N_{\max },\,N_{\min }$ as follows: 
\begin{eqnarray*}
\overline{N} &=&\frac{(N_{\max }+N_{\min })}2, \\
\alpha &=&\frac{(N_{\max }-N_{\min })}{(N_{\max }+N_{\min })}.
\end{eqnarray*}
Given a time $T_{fp}$ between the two oscillating field pulses, if the
oscillating field is applied for a time $t$ at both the beginning and the
end of the storage time then the linewidth $\Delta \nu $ is given by \cite{may_phd} 
\begin{eqnarray}
{\Delta \nu } &=&{{\frac 1{2(T_{fp}+4t/\pi )}}}  \label{eqn: linewidth} \\
&\approx &\frac 1{2T_{fp}}, \,\,{\rm if\, 4t/\pi \ll T_{fp}.}
\end{eqnarray}

\eqnabbr~(\ref{eqn: ramsey lineshape}) may be differentiated to obtain 
\begin{equation}
{\frac{dN}{d\nu }}={\frac \pi {\Delta \nu }}\alpha \overline{N}\sin \left( {%
\frac{\pi (\nu -\nu _0)}{\Delta \nu }}\right) .
\end{equation}
The measurements were made at $\nu \approx \nu _0\pm \Delta \nu /2$, where
the number of neutrons counted was $N_{\uparrow \downarrow }\approx \overline{%
N}_{\uparrow \downarrow }$ for each spin state, giving a total of $N\approx 
\overline{N}_{\uparrow }+\overline{N}_{\downarrow }$ neutrons per
measurement cycle. The fractional uncertainty in the number of neutrons
counted is at best $1/\sqrt{N}$, so the uncertainty in the measurement of
the frequency is no better than 
\begin{eqnarray}
{\sigma _\nu } &=&{{\frac{\Delta \nu }{\pi \alpha \sqrt{N}}}}  \nonumber \\
&\approx &\frac 1{2\pi \alpha T_{fp}\sqrt{N}}.
\end{eqnarray}

In the case of a perfectly constant magnetic field, the EDM could be
calculated from the difference in precession frequency between the two
directions of the electric field. For a total (over a number of measurement
cycles) of $N_T$ neutrons, equally divided between the two directions of the
electric field, the uncertainty in the EDM due to neutron counting
statistics would be 
\begin{equation}
\sigma _d\approx {\frac \hbar {2\alpha E_0T_{fp}\sqrt{N_T}}}.
\label{eqn:edm_stat_err}
\end{equation}
This result, which is applicable when the noise does not exceed that due to
normal counting statistics, corresponds to the fundamental limit of
sensitivity given by Heisenberg's uncertainty principle: the uncertainty in
frequency is inversely proportional to the observation time $T_{fp}$.

It is desirable for the systematic error in absolute frequency to be as
low as 0.2 ppm. In the neutron case there is a significant 
upward shift created by the Ramsey-Bloch-Siegert (RBS) effect \cite{ramsey55,bloch40}. In the EDM data taking cycle, this shift is calculated to be 0.15 ppm. Other systematic-error frequency shifts, such as that due to the rotation of the Earth, are discussed in \cite{baker06} and \cite{baker07}. 

One of the great virtues of the 
Ramsey method is the symmetry of the central fringe about 
the true Larmor frequency (plus RBS shift), even when the fringes are smeared by field inhomogeneities.  In this experiment the Ramsey pattern contained about 100 fringes, and the field was homogeneous to 0.1\%.

\begin{figure} [ht]
\begin{center}
\resizebox{0.5\textwidth}{!}{
\includegraphics{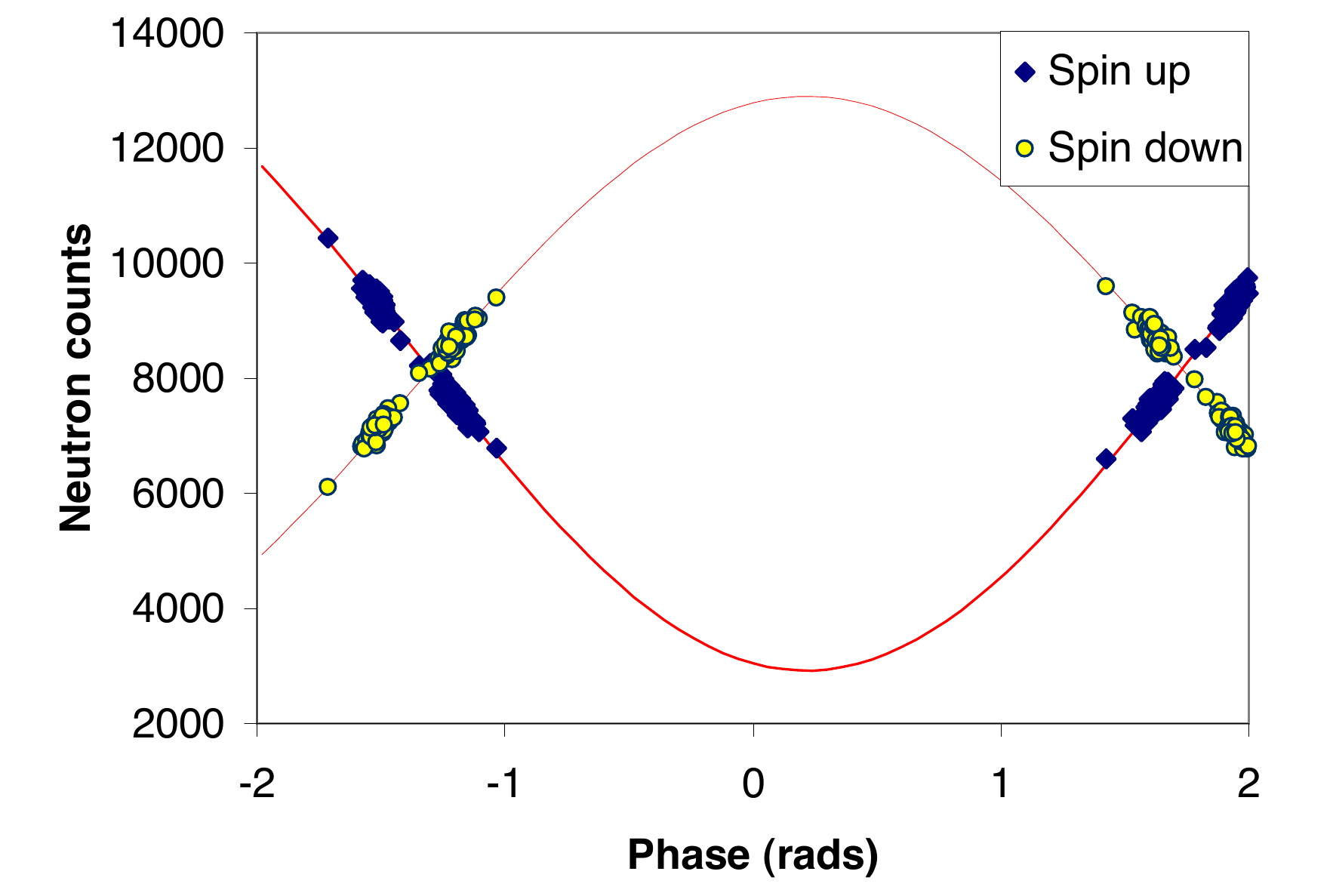}}
\end{center}
\caption{(Color online) Spin-up and spin-down neutron counts for a single run fitted to the Ramsey curve (\eqnabbr \ref{eqn: ramsey lineshape}).}
\label{fig:ramsey_fit}
\end{figure}

Under normal running conditions, the magnetic field drifted slowly.   However, the frequency measurements of the mercury magnetometer allowed  us 
to set up a neutron resonance frequency on the synthesizer unfailingly extremely close to
the desired part of the central fringe, and thereby to compensate for the
magnetically induced frequency shifts within
each measurement cycle.  The  precision of the Hg magnetometer was sufficient for the 
uncertainty on $d_{n}$ to be dominated by neutron counting
statistics, such that equation (\ref{eqn:edm_stat_err}) still applies.  \figabbr \ref{fig:ramsey_fit} shows a typical set of data from a single run, fitted to the Ramsey curve.  The spread of points along the curve arises from the shifts in the magnetic field from one batch cycle to another.

\begin{figure} [ht]
\begin{center}
\resizebox{0.5\textwidth}{!}{
\includegraphics{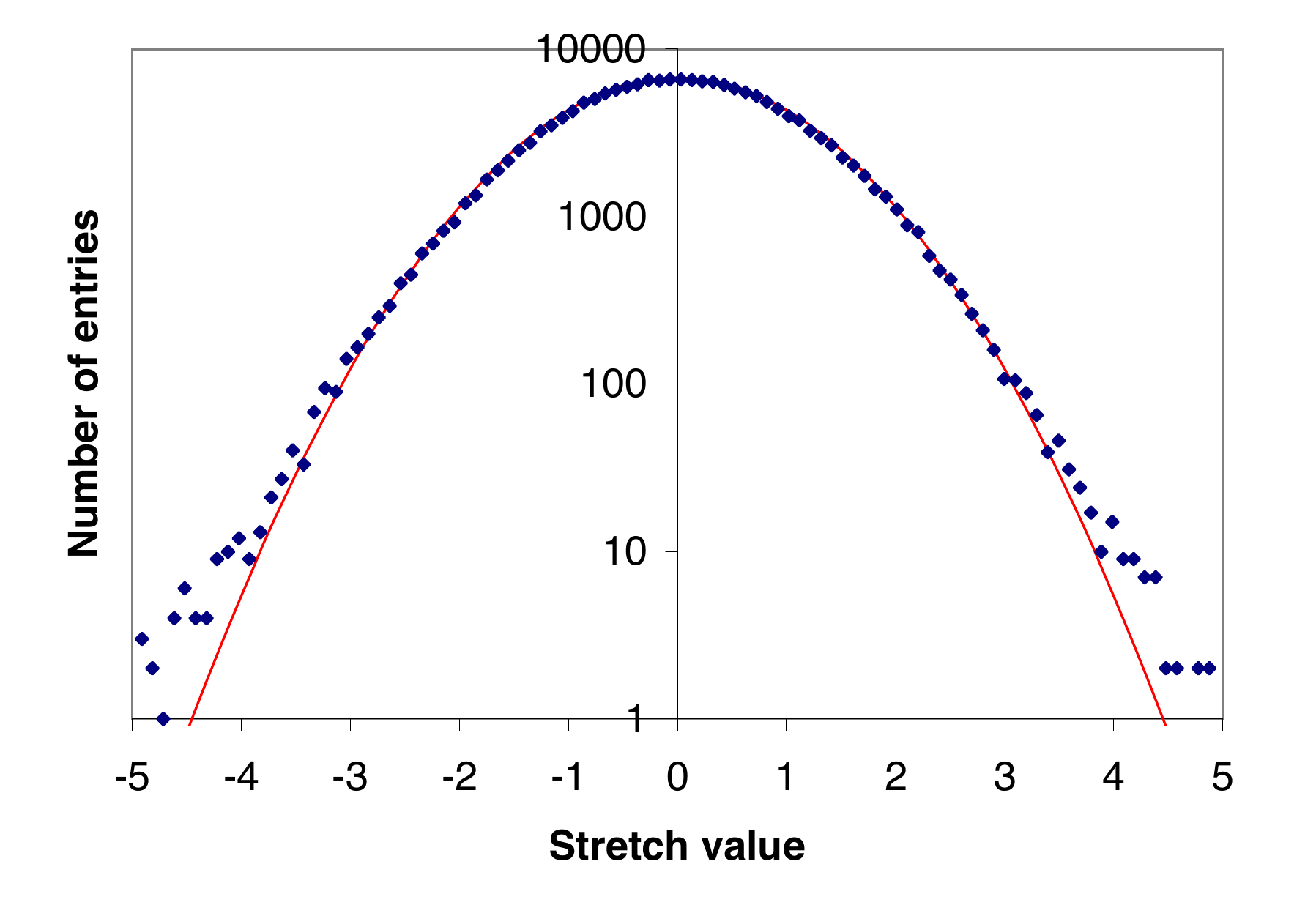}}
\end{center}
\caption{(Color online)  Distribution of stretch values from the fits to the Ramsey curve }
\label{fig:ramsey_residuals}
\end{figure}

The data points of \figabbr \ref{fig:ramsey_residuals} show, on a log scale, the distribution (over the entire data set) of stretch values $r_i$ of the fits to the Ramsey curve: 
\begin{equation}
\label{eqn:ramseystretch}
r_i = \frac{\left(\nu_i -\nu_{R_i}\right)}{\sigma_i},
\end{equation} 
where $\nu_i$ is the calculated frequency of the $i$th batch of neutrons, $\sigma_i$ is its uncertainty and $\nu_{R_i}$ is the expected frequency for that batch as determined by the mercury magnetometer, the applied r.f.\ and the Ramsey curve function.  Ideally, and in the absence of any EDM-like signals, this distribution would be expected to be a Gaussian of unit width.  The continuous line is a Gaussian of width 1.06.   The true distribution departs from this Gaussian at about 4$\sigma$.  The few points lying outside this range tend to be associated with runs that have other known problems, for example with intermittent failure of the neutron delivery system.  Because of the symmetric way in which the data were taken, rejecting batches that lie within the tails from this distribution cannot of itself induce a false EDM signal.

\section{Experimental apparatus}

\begin{figure} [ht]
\begin{center}
 \resizebox*{0.5\textwidth}{!}{\includegraphics[clip=true, viewport = 70 265 484 754]
{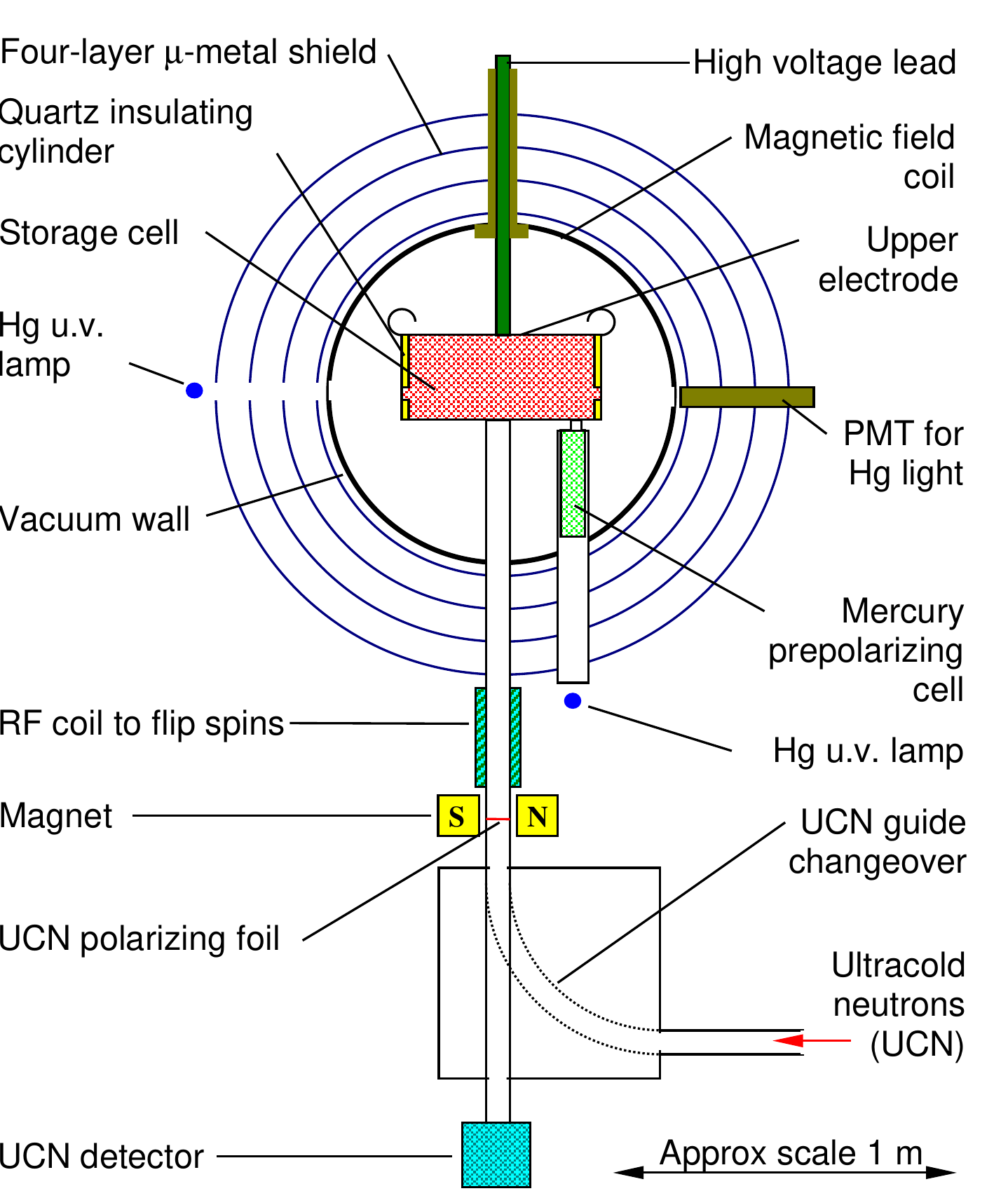}}
\end{center}
\caption{(Color online)  The neutron EDM experimental apparatus}
\label{fig:exptl_apparatus}
\end{figure}

A schematic of the experimental apparatus is shown in \figabbr\ref{fig:exptl_apparatus}. 

\subsection{The neutron subsystem}


\subsubsection{Neutron production and transport}

      Very cold neutrons with a speed of about 50 m/s are extracted from the liquid-deuterium cold source of the 58 MW high-flux ILL reactor, through a vertical guide known as the TGV ({\it tube guide verticale}).  These neutrons are  incident on the Steyerl turbine\cite{steyerl86,steyerl89} which converts them to UCN by reflection from the (receding) turbine blades.  The UCN exiting from the turbine can be directed to several experimental positions by computer-controlled switching of horizontal UCN guides.  At the entrance to the horizontal guide of the EDM position, the turbine blades produce a phase space density (PSD) of 0.084 UCN {(m/s)$^{-3}$cm$^{-3}$}, which remains constant up to a UCN velocity of 8 m/s, or an energy equivalent to 3.2 m fall in height. This PSD can provide 87 UCN/cm$^3$ in a natural nickel bottle, 71 UCN/cm$^3$ in a stainless steel bottle and 25 UCN/cm$^3$ in a vitreous quartz bottle of height 0.12 m. The latter density is the most relevant since the sidewall of the EDM measurement bottle was made of vitreous quartz and was 0.12 m high. These numbers are `real UCN'\cite{steyerl89} in that they do not allow for the reduction on conversion to counts caused by the efficiency 0.80$\pm0.05$ of the UCN detector.
 
      The UCN guide from the turbine blades to the EDM bottle had a total length of 9.2 m divided into a horizontal length of 7.1 m followed by a vertical length to the upper surface of the lower electrode of 2.08$\pm 0.05$ m. This latter figure is the height above the UCN source of the DLC surface of the bottle lower electrode. Thus, UCN need to have an energy corresponding to 2.08 m of height at the source in order to only just reach the lower electrode surface, and to have an energy corresponding to 3.00 m of height at the source in order to enter the EDM bottle with the highest fully containable energy of 0.92 m at the lower electrode surface. This range of energies at the source is within the range of its uniform brightness. Thus a perfect 9.2 m of guide with no polarizer and no safety window in place, and no annihilation of UCN, would fill the EDM bottle to 25 UCN/cm$^3$. 

     We have used diffusion theory\cite{golub79} to model the filling of the bottle with the real guides and their losses. The guides had three types of surface: natural nickel evaporated onto thin glass for 1.8 m in the turbine house, with cross section 70 mm x 70 mm; $^{58}$NiMo sputtered onto electro-polished stainless steel surfaces for 5.9 m from the turbine house to the position of the polarizer, with circular cross section of diameter 78 mm; and Be sputtered onto glass for the 1.5 m above the polarizer to the EDM bottle, again circular in cross section, with a diameter of 65 mm. The theory indicates that at the completion of a long filling of the EDM bottle, the guide system, including the 0.1 mm thick aluminium safety window, is attenuating the PSD at the base of the bottle relative to that of the UCN source by a factor of 0.55 for the lowest energy UCN that can enter the bottle and by a factor of 0.22 for the highest energy UCN that can be contained in the bottle. This represents a considerable softening in the UCN spectrum in the bottle compared to a Maxwell spectrum with the quartz cut-off. There are three mechanisms involved in this softening. First, the UCN that can only just enter the bottle are on the point of marginally exceeding the lower (2.0 m) Fermi potential energy in the (ferro-magnetic) nickel surface of  the guide in the turbine house. This energy excess increases to 1.0 m height equivalent at the top end of the bottle spectrum, and causes much leakage of these UCN through the nickel guide wall. The result is a 30\% relative reduction in the UCN PSD at the top end relative to the bottom end of the bottle spectrum. Secondly, the performance of the entire guide system deteriorates with increasing UCN energy since both the UCN losses in guide wall reflections and their diffuse reflection probabilities increase with UCN energy. This results in a further 29\% relative reduction in PSD at the top end.  Lastly, the UCN current drawn from the guide by the UCN losses in the EDM bottle itself also increases with UCN energy, causing a relative reduction of 17.5\%. When the polarizer is inserted, these last attenuations are slightly more than those just given. 

    The diffusion model just referred to has just one adjustable parameter, which represents the probability of diffuse reflection per collision for UCN with a total energy equal to the critical energy.  All of the guide surfaces have thin sputtered or evaporated coatings on highly polished substrates. The parameter was adjusted to give the observed number of UCN just after filling for our EDM bottle after five filling time constants. Agreement with experiment on UCN densities was therefore ensured. The probability of diffuse reflection per collision deduced from the fit for UCN at the local critical energy was found to be 0.075. In independent experiments, we have found a corresponding value of 0.040 for uncoated lightly electro-polished honed stainless steel surfaces. \cite{taylor_phd, alayoubi_phd} This suggests that coating processes increase the surface roughness for the surface wavelengths that are short enough to produce totally diffuse reflections. For UCN with an isotropic distribution of velocities and kinetic energy equal to half of the critical energy our probability of diffuse reflection per collision on the coated surfaces would be 0.075/2 = 0.038. In the case of uncoated stainless steel this last figure would be 0.02.
 
     The main value of the diffusion calculation has been the determination of the shape of the UCN energy spectrum used for the EDM measurement. The spectrum shape is important in understanding some of the later results. Although the softening of the spectrum reduces UCN numbers, it increases the average UCN storage time more than in proportion to the reduction of UCN energy. This largely cancels the reduction in sensitivity of the EDM measurement by allowing the use of a longer Ramsey resonance time.  The softening also increases the average height difference due to gravity between the stored UCN and the stored Hg atoms.  Knowledge of the UCN spectrum allows one to calculate this height difference, which is needed for a method of assessing the systematic errors caused by geometric phases.\cite{pendlebury04} This height difference can also be determined using magnetic resonance, with a containment trap of variable height.  This gives results in good agreement with that calculated from the UCN spectrum. This UCN spectrum is also successful in fitting the observed UCN counts versus storage interval for all intervals between 60 s and 600 s to within the RMS noise of about 2\% arising from fluctuations in shutter timing. At zero containment time there appears to be a 25\% UCN excess due to the presence of UCN that are not fully contained. At a containment time of 60 s these extra UCN appear to have fallen below the 2\% noise level.

    The spectrum-weighted average attenuation of the PSD in the EDM bottle filling process was a factor of 0.295 relative to the UCN source. This led to an initial density of fully contained UCN in the bottle, after a long filling time with no polarizer and on just closing the door, of 7.5 fully contained UCN/cm$^3$ and a total number of 160,000 UCN. The latter number falls to 69,400 after the containment interval of 140 s used when taking EDM data. To find the final UCN counts from an EDM data-taking cycle we must take account of further attenuations to the figure of 69,400 per batch. These are (i) 0.727 for curtailment of the filling and emptying intervals to conserve polarisation and batch cycle duration (ii) 0.525 for spin selection, which includes a small increase due to production of wrong spins (iii) 0.80 for the combined loss in two transits of the polarizer foil (iv) 0.875 for losses when waiting for the spin flipper while the other spin state is counted (v) 0.915 for guide losses in transit from the bottle to the detector (vi) 0.80 for detector efficiency. These figures indicate a final count of 13,600 per batch - close to the 14,300 observed average count from all runs.
 
   We believe that the spectrum changes derived from these last attenuations are small and partly cancelling - process (i) gives a slight hardening (ii) and (iii) and (vi) are neutral while (iv) and (v) induce a slight softening.

In order to deal with the variety of surfaces involved, a simple model has been adopted for estimating the parameter $\eta$ to be used with the theoretical energy dependence in calculating the UCN loss probability per collision.  Our model takes $\eta=( \eta_A + \eta_H)$, where $\eta_A$ is the contribution for the atomic composition of the material excluding hydrogen and $\eta_H$ is the contribution from interstitial hydrogen. We are concerned with the situation where none of the materials has been baked in vacuum. From measurements on 316-type stainless steel\cite{mampe80} we take $\eta_H$(SS) to be $3.9\times10^{-4}$  and for other materials X we take $\eta_H$(X)  = $\eta_H$(SS)$\times (V_{SS}/V_X)$, where the $V$s are the mean Fermi potentials. This amounts to assuming that, at room temperature, the atomic fraction of hydrogen and the UCN loss cross-section for hydrogen are the same in the surface layers of all the materials concerned. In our experience this model works well in predicting lifetimes to about 20\% in wide variety of bottles and guide tubes made of unbaked materials at room temperature.

The key data used in this assessment arose from a data-taking run labelled ALP1120.dat, which produced data for UCN counts versus containment time in steps of 5 s up to 660 s. It used a large  smooth-sided bottle with a period of 60 s used for filling and 70 s for emptying, with no polarizer present. Only the emptying process enters to cause the data count totals to differ from the actual number of real UCN in the bottle when the shutter is opened for emptying. This difference involves just two factors: (i) the detector efficiency, and (ii) UCN lost in the emptying guide and in the bottle after the shutter is opened. The detector efficiency is generally assessed as 0.80$\pm0.05$, with losses in the window and loss of counts below the discrimination level each being about 0.10. The overall emptying time constant after 140 s of containment was measured in a separate run, labelled ALP1115.dat, to be 9.35$\pm0.30$ s. After this containment, the bottle UCN lifetime is about 210 s, so the fractional bottle loss during emptying is to first order 9.4/210 = 0.045. To calculate losses in the guide we need the average time spent in the guide by each UCN before it is detected, and the storage time of the guide. The latter is typically 20 s. If the guide were to be perfectly smooth the time to the detector would be the free-fall time, which is 0.4 s; however, the guide has some roughness, and we can estimate from the emptying time constant that about 40 \% of the UCN that leave actually return to the bottle. Assuming that the roughness approximately doubles the time taken, making 0.8 secs, the fractional loss would be 0.8/20 = 0.04, making a total emptying loss of 0.045+0.04 = 0.085. We are now in a position to calculate the real number of stored UCN. Then, knowing the turbine performance\cite{steyerl86,steyerl89}, we have the overall loss in the entry guide system, which allows us to fix the roughness parameter.

\subsubsection{The neutron polarizer}

The neutrons were polarized by transmission through a silicon foil upon
which was deposited a $1~\mu$m layer of iron that was magnetised close to
saturation by a field of about 0.1 T from a permanent magnet. This had Fermi
potentials of approximately 90 and 300 neV for the two spin states of the
neutron. The foil was mounted 1.5 m below the trap, so that neutrons that
had sufficient energy to penetrate the foil could slow down before reaching
the trap.

The polarizer was mounted with the magnetized layer towards the trap, since
experience in the past showed that this orientation gave the better
neutron polarization. With neutrons that made a single passage through the
foil, such polarizers can produce a transmitted neutron polarization in
excess of 90\% \cite{herdin77}. However, they do have a
finite probablity, of a few percent, of flipping the spin of both
transmitted and reflected neutrons. In this case it led to a build-up of neutrons in
the unwanted spin state as the trap filled, thus reducing the polarization that
was finally achieved. The maximum polarization was obtained for very short
filling times \cite{miranda_phd}.  The filling time was therefore adjusted so as
to maximize $\alpha \sqrt{N}$.

As mentioned above, the 1.5~m of neutron guide between the polarizer and the
neutron trap was made of glass, with the inner surface coated with BeO, 
which is non-magnetic. This guide was used instead of a stainless steel
guide because remnant magnetization and magnetic domain structure in a
stainless steel guide would have caused severe inhomogeneity in the $B_{0}$ field
as well as causing depolarization of the neutrons in wall collisions. The
use of glass also allowed the penetration of the oscillating magnetic field
of the spin flip coil, at 20~kHz. This coil was used towards the end of the
measurement cycle, when the spin-down neutrons were emptied from the
trap and counted.

To prevent depolarization as the neutrons passed from the magnetic field of
the polarizer, through the Earth's 60 $\mu$T magnetic field, and into the 1 $\mu$T magnetic
field of the trap, a variable-pitch solenoid was wound around an 18-cm-diameter former concentric with the
guide tube. This ensured that the magnetic field changed smoothly and
monotonically, and that there is no zero-field region along the guide.

\subsubsection{The neutron storage trap}

The neutron storage trap was made of two flat, 30 mm thick, circular aluminum electrodes, 
separated by a hollow right circular cylinder of quartz that also acted as
a high-voltage insulator. The electrodes had aluminum corona domes
attached, and the insulator was recessed 15 mm into the electrodes to reduce
high-voltage breakdown \cite{alston}. At the bottom of the recess in
each electrode, a Teflon O-ring was housed to provide a gas-tight seal
between the electrode itself and the inner surface of the quartz ring, so as
to contain the polarized atomic mercury used for the magnetometry, as described in \secabbr \ref{sec: magnetometer}.

About halfway through the data-taking period, the existing smooth-walled quartz cylinder was replaced by another quartz cylinder of the same inner dimensions but with a matt surface finish.  These are referred to as the smooth and rough traps respectively.

Bare aluminum has a Fermi potential of 55 neV (corresponding to a
critical velocity of  3.3 m/s).  Aluminum oxide surfaces quickly 
depolarize any mercury that comes into contact with them. The electrodes are
therefore coated with a thin insulating layer of a relatively high Fermi
potential material. Initially, Teflon was used for this purpose; it was
sprayed on, and baked in an oven. However, it did not adhere well enough to
the surface, and it eventually peeled away, causing high-voltage sparks to the resulting loose Teflon flaps. The Teflon was then
replaced by a 1 $\mu $m thick coating of diamond-like
carbon (DLC), produced by chemical vapor deposition from a plasma discharge
in deuterated methane \cite{grinten99}, which proved to
be far more durable. The Fermi potential of this layer is 220 neV. The
quartz insulator has a Fermi potential of 91 neV.  All of the data analysed in this paper were taken with the DLC-coated electrodes.

The trap had an interior diameter of 470~mm and a height of 150~mm. The 15
mm recess in each electrode yielded a distance between the electrodes, for
the majority of the surface, of 120 mm. The overall volume was therefore
21~liters.

The annular quartz insulators forming the sidewalls, which
were machined from single pieces of fused silica, had a 15 mm wall thickness. 
A Suprasil window in
either side allowed the passage of a beam of polarized 2537 \AA\ light, which
was used to probe the state of polarization of the mercury atoms as they
precessed in the $\vecB$ field.

The lower electrode was electrically grounded, and had a 67~mm diameter, 4 cm deep hole
in the center, through which the neutrons enter the trap. The hole could be
closed by a sliding DLC-coated beryllium-copper door that had been adjusted to
have gaps of less than 100~$\mu $m. This non-magnetic door slid on nylon
bearings and it was operated by a mechanical coupling from a remote piston driven
by compressed air. A second hole in the electrode, of diameter 10
mm, gave access to a door that opened for 1 s during the measurement
cycle to allow the polarized mercury to enter the trap.

The neutron-trap support system, door mechanism, mercury polarizer and all
other items inside the vacuum vessel were made from non-ferromagnetic
materials. Materials such as brass were avoided because they often contain
ferromagnetic impurities.  Scans  with a fluxgate magnetometer of sensitivity 1 nT approaching to within 2 cm of the inner surface of the storage volume revealed no magnetic anomalies. 

\subsubsection{The neutron detector}

The neutron detector was a proportional counter containing 1200 mbar of
argon, 50 mbar of $^{3}$He and 100 mbar of methane, in which the neutrons
were detected via the reaction 
\begin{equation}
n+\,^{3}\mathrm{He}\rightarrow \ ^{3}\mathrm{H}+p,
\end{equation}
which releases 764~keV of energy. The central electrode was a loop of
tungsten wire of diameter $200~\mu $m and was maintained at 2.5~kV \cite{taylor_phd}.

The window of the detector was a 100~$\mu $m aluminum foil, with a mean
Fermi potential of 55~neV. The detector was placed 2~m below the neutron
trap to ensure that nearly all the neutrons reaching it, after falling
freely through the Earth's gravitational field, have a sufficiently large
velocity component perpendicular to the window to penetrate it. The
efficiency of the detector was about 80\% for UCN. The detector was shielded
by 150~mm of polyethylene and 5~mm of boron-loaded plastic resulting in a
background \textit{in situ} of less than one count in 10~s, whereas the average UCN
count after a single four-minute measurement cycle was about 14,000 in 40~s for this data set.

\subsection{The mercury magnetometer}

\label{sec: magnetometer}
\begin{figure} [ht]
\begin{center}
\resizebox{0.5\textwidth}{!}{\includegraphics*{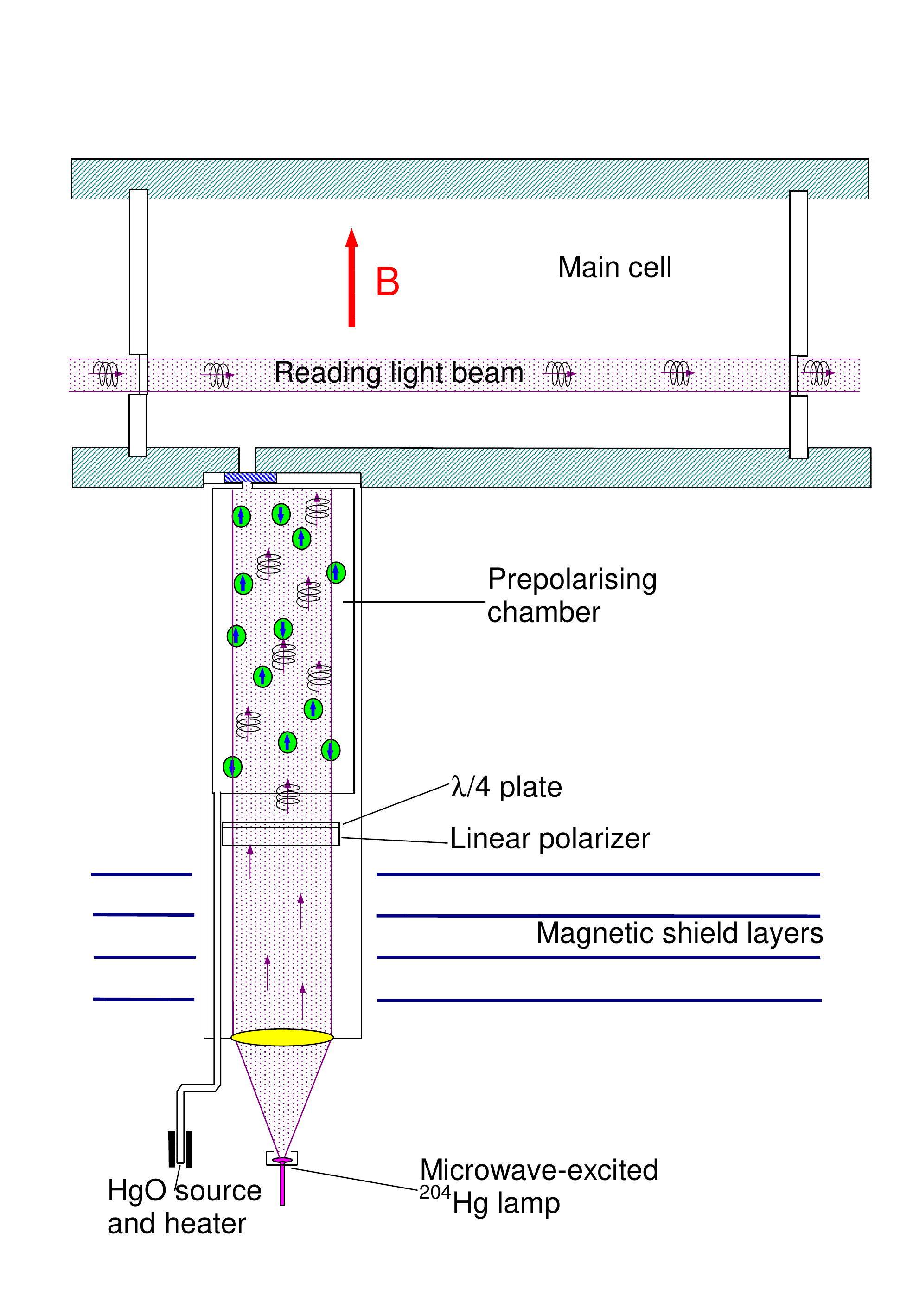}}
\end{center}
\caption{(Color online) The mercury magnetometer}
\label{merc_mag}
\end{figure}

The construction and performance \cite{green98} of the atomic mercury
magnetometer (\figabbr~\ref{merc_mag}) have been discussed elsewhere.
Here a brief account is given of its use in the EDM experiment.

\subsubsection{Principle of operation}

\begin{figure} [ht]
\begin{center}
\resizebox{0.5\textwidth}{!}{
\includegraphics{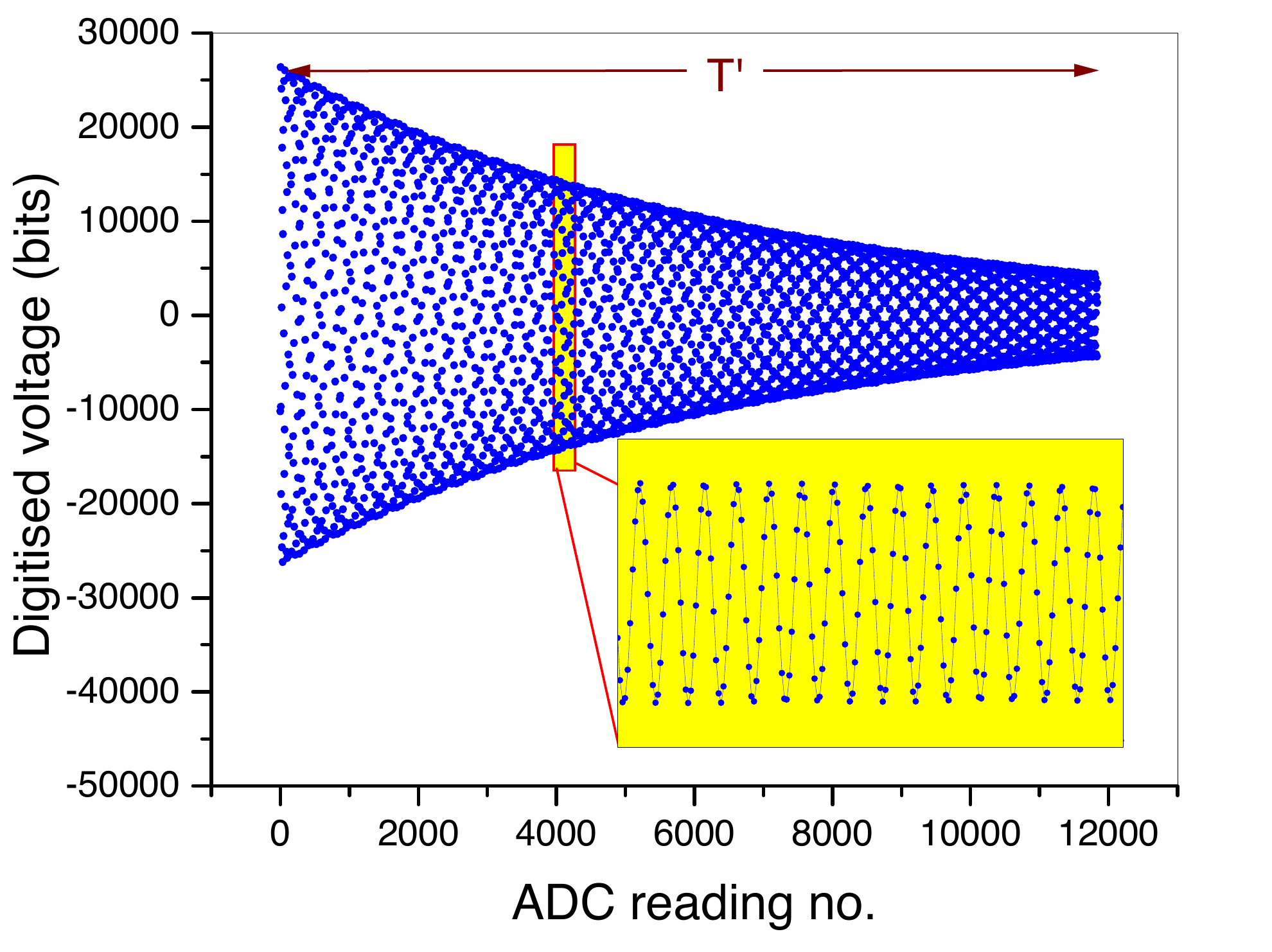}}
\end{center}
\caption{(Color online) A set of mercury ADC readings from one
measurement cycle. The gradual depolarization is clearly visible, and the
expanded region shows the underlying 8 Hz precession frequency.  The frequency measurement period $T^\prime$ excludes a two-second settling period at the start.}
\label{Hg_adcs}
\end{figure}

Spin-polarized $^{199}$Hg atoms were made to enter the storage volume once it
had been filled with neutrons and the neutron entrance door had been closed.
A rotating magnetic field ${\bf B}_1^{\prime }$, perpendicular
to the main $\vecB$ field, was applied  for a period of 2~s. The ${\bf B}_1^{\prime }$
field had a frequency  equal to the spin precession frequency of the mercury
atoms -- 7.79 Hz -- and was of the appropriate strength to turn the spin
polarization vector by $\pi /2$ radians into the $xy$ plane perpendicular
to $\vecB$. Meanwhile, a beam of 2537 \AA\ polarized light from an
isotopically-pure $^{204}$Hg discharge tube traversed the chamber. The
absorption of this light depended upon the $x$ component of polarization of
the mercury atoms, and thus varied with time as an exponentially-decaying
sinusoid. The intensity of the light was monitored by a solar-blind Hamamatsu R431S photomultiplier tube, the output current of which was converted into a voltage and digitised
with a 16-bit ADC at a rate of 100 Hz. The resulting data (\figabbr~\ref{Hg_adcs}) were fitted to
obtain the average frequency, and hence the volume- and time-averaged
magnetic field, during the Ramsey measurement interval. At the end of the storage period
the mercury atoms were pumped out of the cell via the neutron entrance door.

\subsubsection{Mercury source, polarizer and analyzer}

The mercury source was a powder of $^{199}$HgO, which was dissociated by
continuous heating to approximately 200 $^{\circ }$C. After passing through
a narrow Fomblin-grease coated pipe, the mercury atoms reached a 1.2 liter chamber situated adjacent
to the main neutron storage volume, within the 1 $\mu $T $\vecB$ field. There
they were optically pumped by light from a $^{204}$Hg discharge lamp,
identical to that used for monitoring the polarization within the neutron
storage volume. The pumping process was continuous, so that, as each charge
of polarized atoms entered the storage volume for the frequency measurement,
the next charge began to build up and polarize.

The $^{204}$Hg discharge lamp lay one focal length below an 80 mm diameter
f2 lens (which also served as a vacuum window). The parallel beam of light
thus produced passed through a linear polarizer followed by a quarter-wave
plate to produce the necessary circular polarization. The analyzing, or
reading, light followed a similar arrangement.

\subsubsection{Absorption and polarization characteristics}

The absorption $A$ of the reading light, which was proportional to the number
of mercury atoms within the chamber, is defined as 
\begin{equation}
A=\frac{I_{0}-I_{1}}{I_{0}},  \label{eqn:absorption_defn}
\end{equation}
where $I_{0}$ and $I_{1}$ are the DC levels of the reading light measured
just before and just after, respectively, the injection of polarized mercury
into the main storage volume. The initial amplitude $a$ of the oscillating
signal is related to the polarization $P$ as \cite{green98,dehmelt57a,dehmelt57b} 
\begin{equation}
a=I_{1}\left\{ \left( 1-A\right) ^{-P}-1\right\},
\label{eqn:ampl_A_P}
\end{equation}
so the level of polarization may be extracted simply from the absorption
and the fitted signal amplitude. 

The polarization is found to depend
strongly upon $A$ because, in the polarizing chamber, the probability of
absorbing a reemitted photon increases quadratically with the density of mercury.
Secondary, and higher order, reabsorptions increase even more quickly. A
large charge of mercury therefore yields a relatively small polarization.
One finds empirically that 
\begin{equation}
P\approx p_{1}\exp \left( -A\alpha\right),
\label{eqn:P_vs_A}
\end{equation}
where $p_{1}$ and $\alpha$ might typically have values of
around 0.5 and 6 respectively. The function (\ref{eqn:ampl_A_P}) 
is maximized at an absorption of approximately 16\%, and this
therefore provides the optimum signal-to-noise ratio. The temperature of the
mercury source was adjusted periodically in order to try to keep the absorption fairly near this value.

\subsubsection{Calculation of precession frequency}

As with other aspects of the magnetometer, the frequency fitting procedure
has been discussed in some detail in \cite{green98}, and it is
therefore only briefly described here.

The AC component of the mercury signal was amplified so as to match the input
voltage range of the ADC used for its digitisation. The clock pulses that
trigger the ADC readings were gated off while the mercury entered the chamber
and while the $\pi /2$ pulse was applied, and readings for an additional 2~s
after that time were ignored in case they were influenced by transient effects. The
readings were, however, recorded throughout the 20~s neutron filling period,
during which time there was no mercury in the storage cell. This allowed the
evaluation of the rms noise on the signal, from which an estimator of
the uncertainty of each reading in the fit could be deduced.

Because the magnetic field drifted with time, the frequency changed slightly
during the measurement. Therefore, instead of fitting the entire array of
ADC readings to a decaying sinusoid, a pair of shorter ($t=15$~s) intervals at
either end of the Ramsey measurement period were fitted in order to find the phases at
points close to the beginning and the end \cite{pendlebury95}. The total
phase difference (including $2n\pi $ for the complete cycles) divided by the
time gives the average frequency, and hence the time- and volume-averaged
magnetic field for the interval of free precession.

\begin{figure} [ht]
\begin{center}
\resizebox{0.4\textwidth}{!}{\includegraphics*[clip=true, viewport = 40 400 518 750]{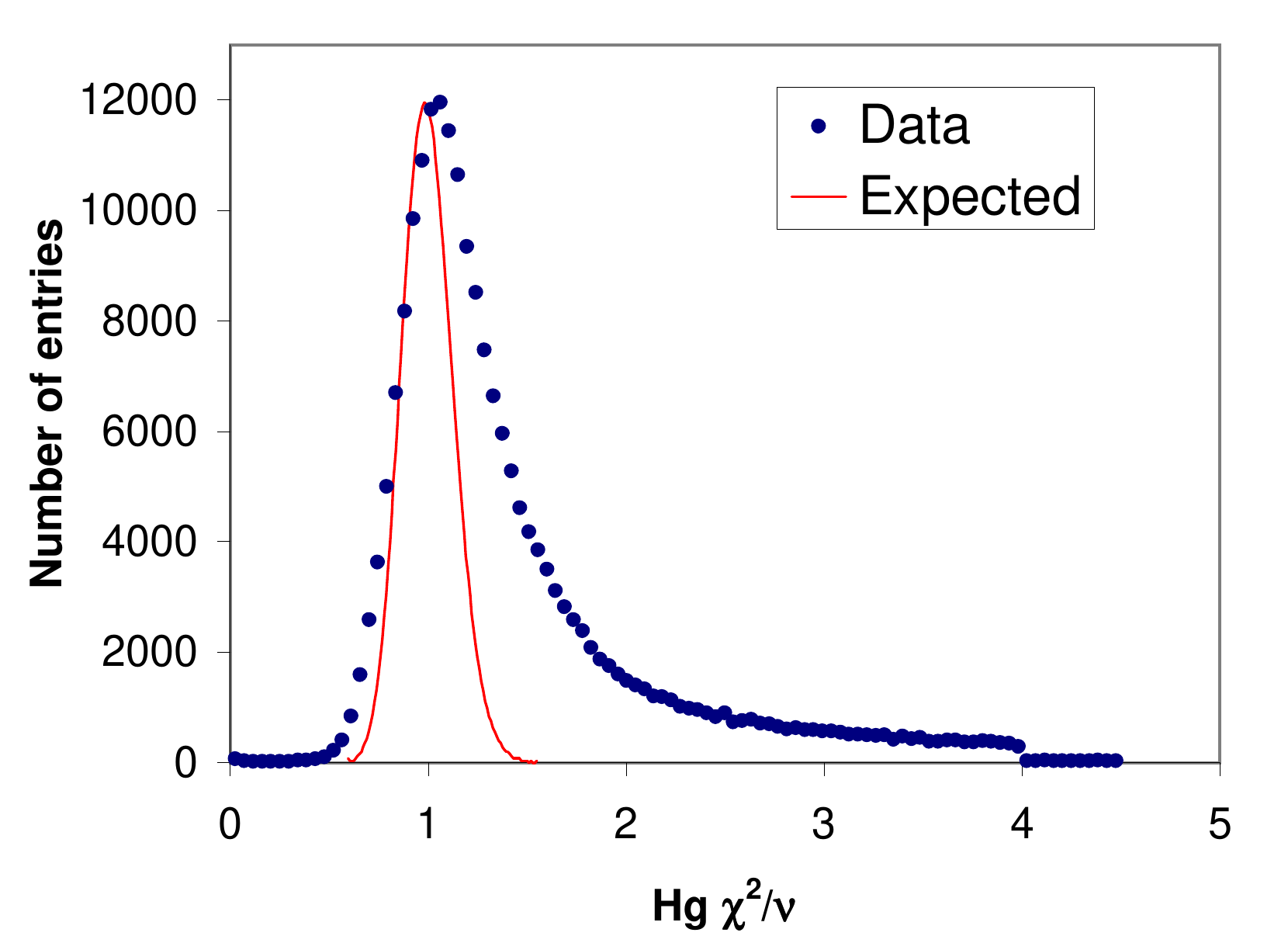}}
\end{center}
\caption{(Color online)  Distribution of $\chi ^2/\nu $
for approximately 205,000 fits of the mercury precession frequency, together with the expected distribution for the ideal case of no magnetic field drift. }
\label{fig:Hg_fit_chisq}
\end{figure}

The fitted function generally
appeared to describe the data well, with the $\chi ^{2}/\nu$ distribution
peaking close to 1.0, as shown by the data points in \figabbr \ref{fig:Hg_fit_chisq}.      

The distribution shown in \figabbr \ref{fig:Hg_fit_chisq} is truncated at $\chinu$ = 4.5.  If $\chinu$ $> 4$, however, the online fitting procedure attempts to correct for potential hardware errors such as missed clock cycles, sticking bits, saturation, too-short depolarisation time, and/or occasional sparks.  The discontinuity at 4.0 reflects the fact that the majority of the fits with originally larger $\chinu$  were incorrect, and they have successfully been re-fitted with an appropriate correction for one or more of these problems.

\subsubsection{Effects of the bandpass amplifier}

\label{sec:bandpass}

The mercury frequency fitting routine assumed no correlations between the individual ADC readings. The measured rms noise was used as an
estimate of the uncertainty of each point. Prior to digitization, however, the mercury signal was
filtered by a bandpass amplifier with a $Q$ of approximately 5.9 in order to
reduce the noise; consequently, neighbouring ADC measurements are actually rather 
strongly correlated with one another, and the calculated variance
must be modified to allow for this.

If the points were independent, the variance $\sigma ^{2}$ of the fitted
frequency would be expected to be inversely proportional to the number of
readings $n = 3000$ obtained in the short intervals at either end of the signal
train, as shown in ref.\ \cite{green98}. When the data are
correlated, this is no longer true; for a given bandwidth, increasing the
sampling frequency beyond a certain point does not reduce the variance. The
calculations in \cite{pendlebury95} suggest that that point is reached
when $n_{s}Q=3$, where $n_{s}$ is the number of readings taken per period.
In the case of this experiment, $n_{s}=12.5$ and $Q\approx 5.9,$ giving an
overall factor of 74, i.e. approximately 25 times above this limit; thus the true
variance on the frequency determination is expected to be higher than the \naive\ estimate by the same factor of 25.  
  
This hypothesis was tested by adding white noise to a precise 8 Hz
synthesized signal from a frequency generator, and performing a series of
fits of the frequency of the resulting signal, firstly with and then without
the bandpass filter in place. With a flat response, the spread in the
measured frequencies was consistent with a Gaussian random distribution
about the mean, having $\chi ^{2}/\nu =1.0$. With the bandpass filter, the
noise was reduced by a factor of five, as was the estimated uncertainty of
each fitted frequency; but the scatter in the results increased, with $\chi
^{2}/\nu $ rising to 25, suggesting that the error bars were indeed a factor
of five too small. Furthermore, this same factor is consistent with the
scatter observed in the experimental data during periods when the magnetic
field is stable, and it also agrees with estimates based upon numerical
simulations using a digital Butterworth filter.

In the discussions that follow, all calculated uncertainties in the
mercury precession frequency incorporate a factor of 5.0 (i.e., a factor of
25 in the variance) to allow for this narrow-banding effect.

This same effect also broadens the $\chinu$ distribution.   The expected distribution, shown as a smooth curve in \figabbr \ref{fig:Hg_fit_chisq}, is therefore that appropriate to 3000/25 = 120 degrees of freedom.  As the magnetic field during each measurement period drifts slightly, the frequency is not perfectly constant.  The true distribution is therefore expected to broaden further, particularly on the high side.  There is a reasonable match on the low side, and the position of the peak is close to unity.

\subsubsection{Performance of the magnetometer}

As with the neutrons, it is desirable that the absolute precision of the mercury frequency measurements should be better than 0.2 ppm.  In \secabbr\ \ref{sec:Hg_accuracy} we discuss possible mechanisms that could affect the accuracy of this system.

\begin{figure} [ht]
\begin{center}
\resizebox{0.5\textwidth}{!}{
\includegraphics{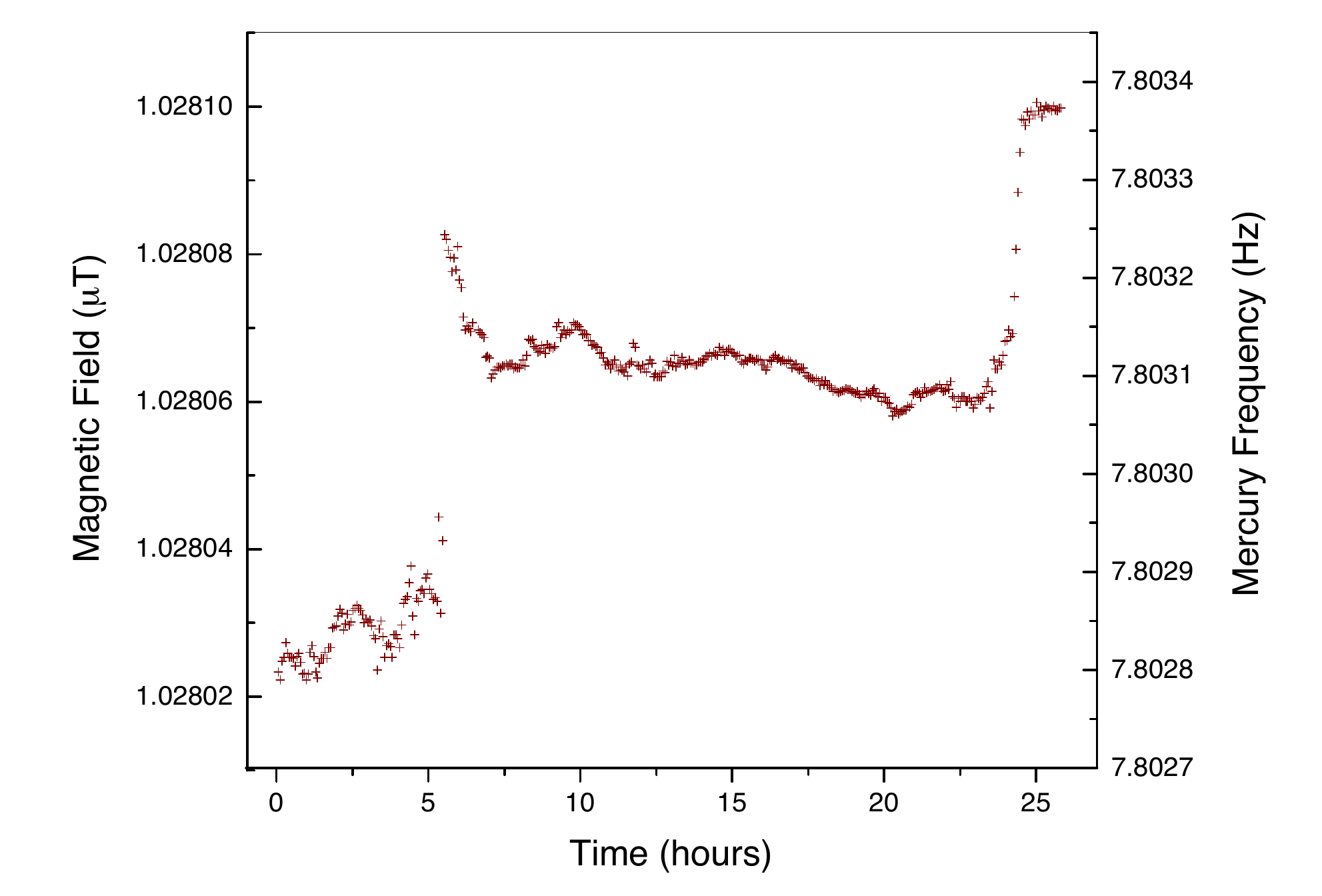}}
\end{center}
\caption{(Color online) Magnetic field strength, as determined by the mercury resonant frequency, measured
repeatedly over a 26-hour period.}
\label{Hg_frequency}
\end{figure}

\figabbr     \ref{Hg_frequency} shows a typical example of the evolution of the magnetic field, 
as measured by the mercury precession
frequency, throughout a typical run. Error bars, which
are of the order of a microhertz, are smaller than the points themselves on this plot.
The drift in magnetic field during this time is approximately $5\times
10^{-11}$ T. For this run an electric field of magnitude 4 kV/cm was applied to the
storage volume, with its polarity reversing approximately every 70 minutes.

\begin{figure} [ht]
\begin{center}
\resizebox{0.5\textwidth}{!}{
\includegraphics{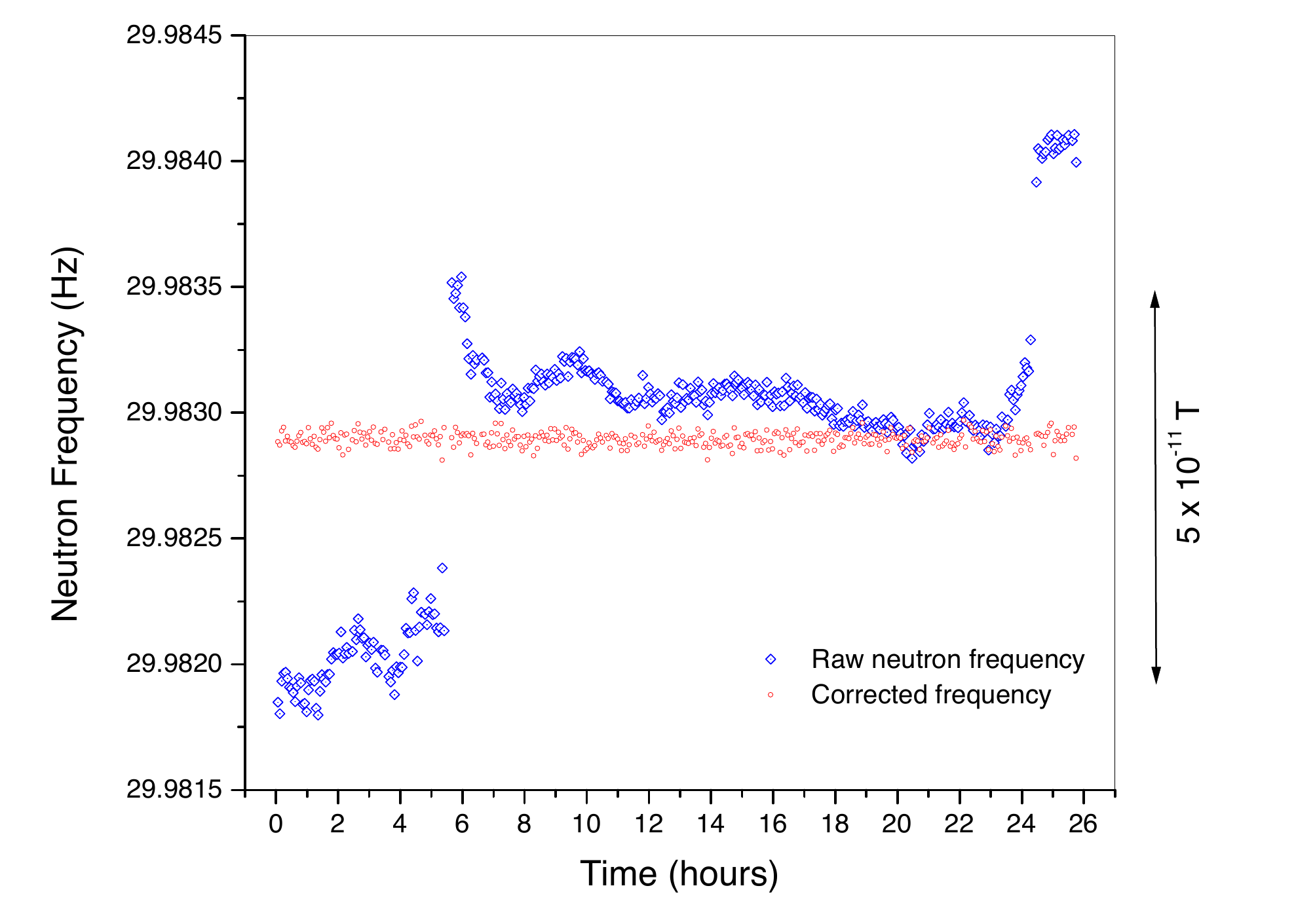}}
\end{center}
\caption{(Color online) Neutron resonant frequency, measured
over the same 26-hour period, before and after correction of the effect of
the drifting magnetic field by normalisation to the measurements of the
mercury magnetometer.}
\label{n_corr_freq}
\end{figure}

\figabbr   \ref{n_corr_freq} shows the corresponding series of
measurements of the neutron resonant frequency throughout the same 26-hour
period.  As expected, the same drift in magnetic field is reflected in
this set of data. Error bars are again omitted for clarity, but are of order
29 $\mu $Hz for this particular data set. The ratio of neutron to mercury
frequencies, normalised to the mean neutron frequency --- i.e., the measured
neutron frequency corrected for the magnetic field drift --- is shown on the
same plot, where it appears as a flat line. The uncertainty on each point is
approximately one part per million, giving a $\chi ^{2}/\nu $ of 0.89; this
is consistent with the width of the line being entirely dominated by
neutron counting statistics. 
Any change in the
neutron resonant frequency due to the interaction of the electric field with
the neutron EDM would appear as a change in this ratio of frequencies. A
straight-line fit to the ratio as a function of the applied electric field
therefore yields a slope that is directly proportional to the EDM signal. It
is evident that the use of this magnetometer compensates extremely efficiently 
for the large-scale effects of magnetic-field drift.

\subsubsection{Mercury frequency uncertainty}

\begin{figure} [ht]
\begin{center}
\resizebox{0.5\textwidth}{!}{\includegraphics*{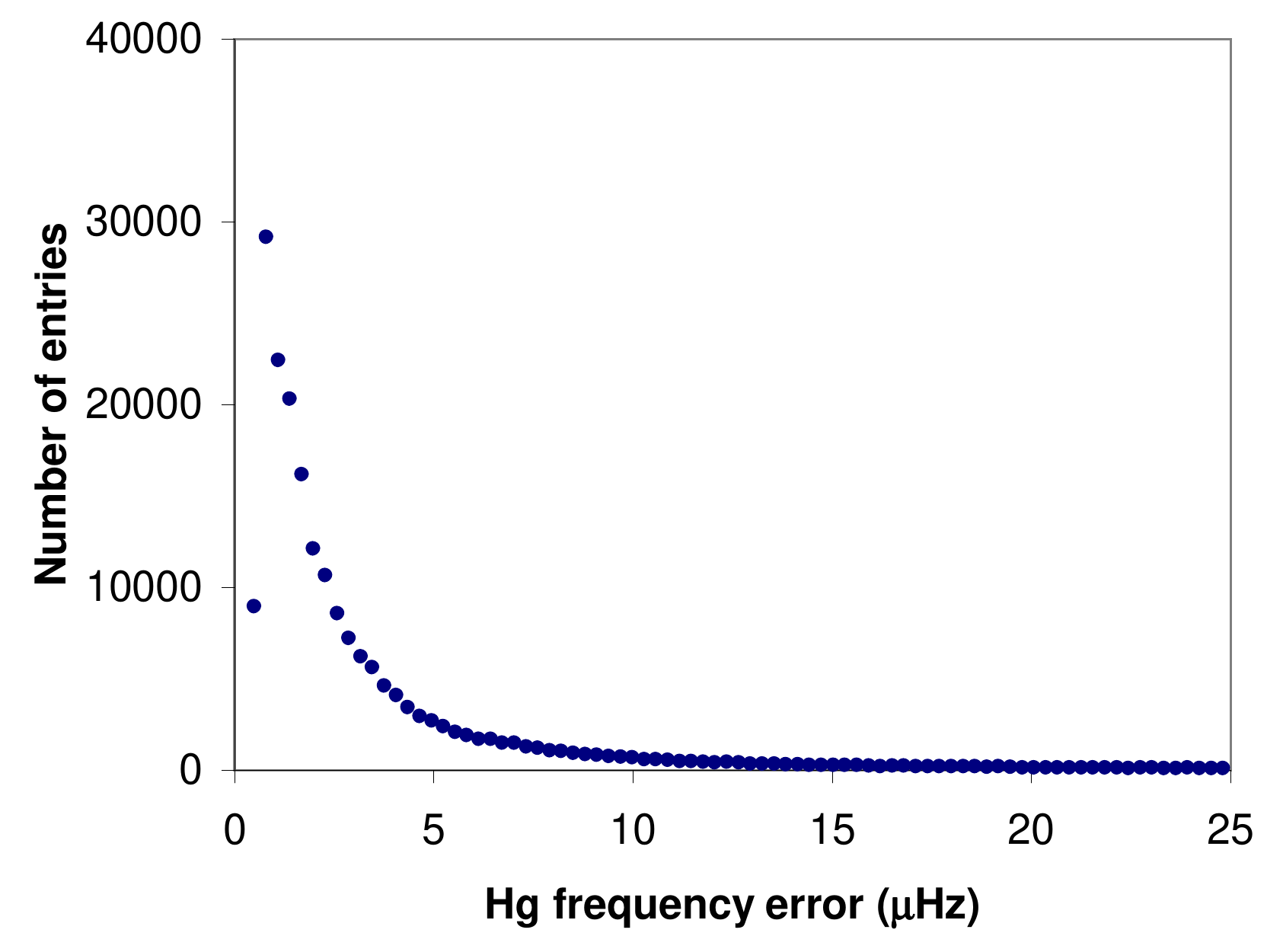}}
\end{center}
\caption{(Color online) Distribution of uncertainties of the fitted Hg precession frequency}
\label{fig:Hg_fit_err}
\end{figure}

The fitted Hg frequency sometimes has a relatively large uncertainty, particularly if the depolarization time is short.  The distribution of these uncertainties is shown in \figabbr \ref{fig:Hg_fit_err}; a typical value is 1-2 $\mu$Hz.   For comparison, the typical inherent neutron frequency uncertainty from counting statistics was about 20  $\mu$Hz, corresponding to about 5 $\mu$Hz in the mercury system.

\subsubsection{Magnetic field jumps}

\begin{figure} [ht]
\begin{center}
\resizebox{0.5\textwidth}{!}{
\includegraphics{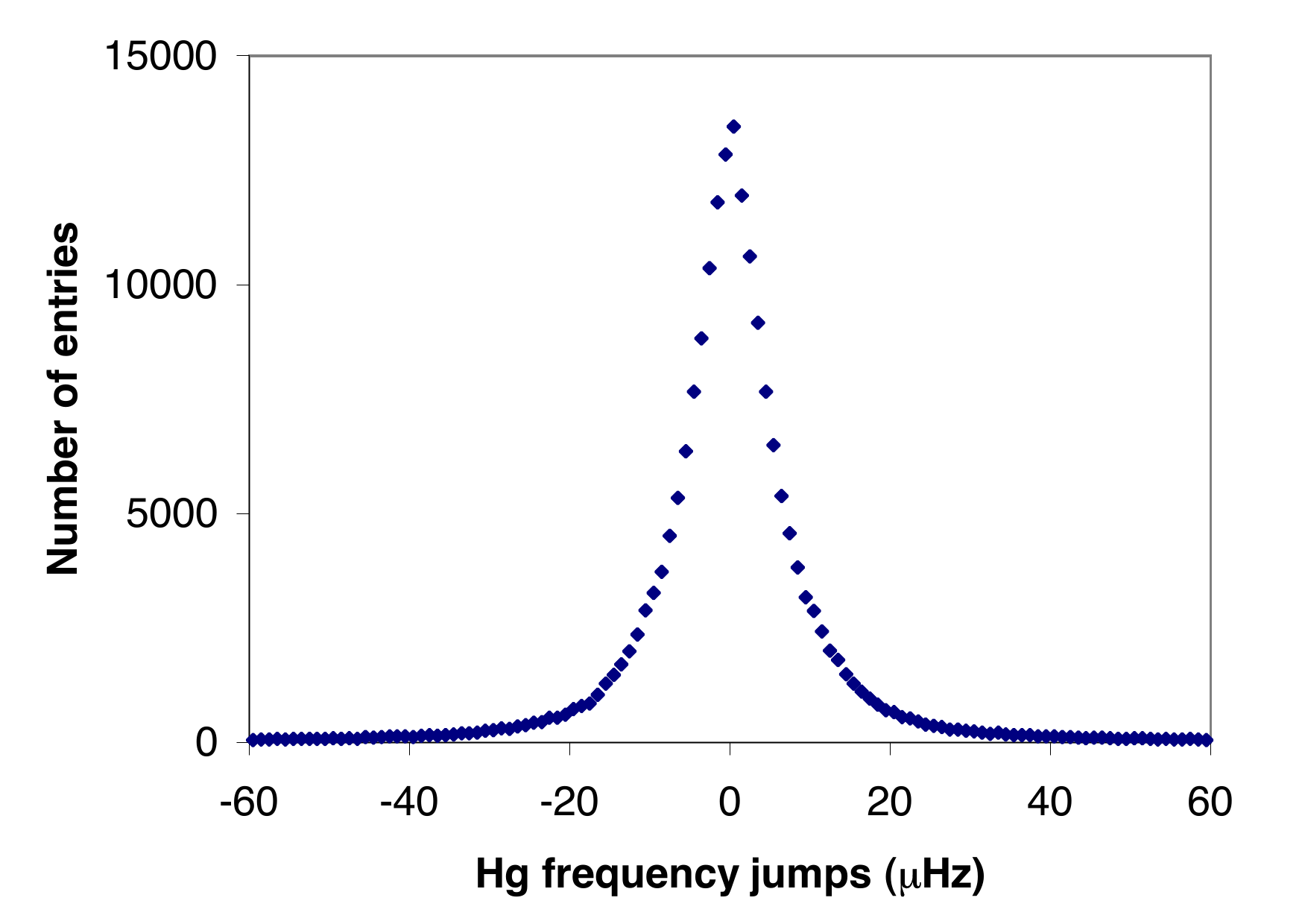}}
\end{center}
\caption{(Color online)  Distribution of changes in the Hg frequency from one batch cycle to the next}
\label{fig:Hg_freq_jumps}
\end{figure}

The distribution of Hg frequency jumps, i.e.\ the difference in Hg frequency between a given batch and the previous batch, is shown 
in \figabbr \ref{fig:Hg_freq_jumps}.  There are broad tails due to occasional sudden changes in field, for example due to the movement of an overhead crane or to a mechanical disturbance to the $\mu$-metal shields.  

The mercury and the neutron frequency measurements do not have perfect temporal overlap.  One can consider the start and end of the Hg measurements to be centred on the 15-second averaging period at the start and end of the Ramsey measurement time, whereas the neutrons average over all but 2 seconds at either end.  If the field is changing, there is therefore a roughly 7-second period - i.e. about 1/30 of the total batch period - for which the change is not properly accounted.  For comparison, a frequency jump of  $60$ $\mu$ Hz -- which would be regarded as extreme --  corresponds to a field jump of about 7.5 ppm, or just over 1/20 of the Ramsey linewidth.  With the aforementioned protection factor of 1/30, this corresponds to a potential error in the frequency ratio $R$ of 0.25 ppm, to be compared with a typical statistical uncertainty on the neutron frequency of about 0.7 ppm.

\subsubsection{Depolarization in strong electric fields}

\label{section: Hg and high voltage}
\begin{figure} [ht]
\begin{center}
\resizebox{0.5\textwidth}{!}{
\includegraphics{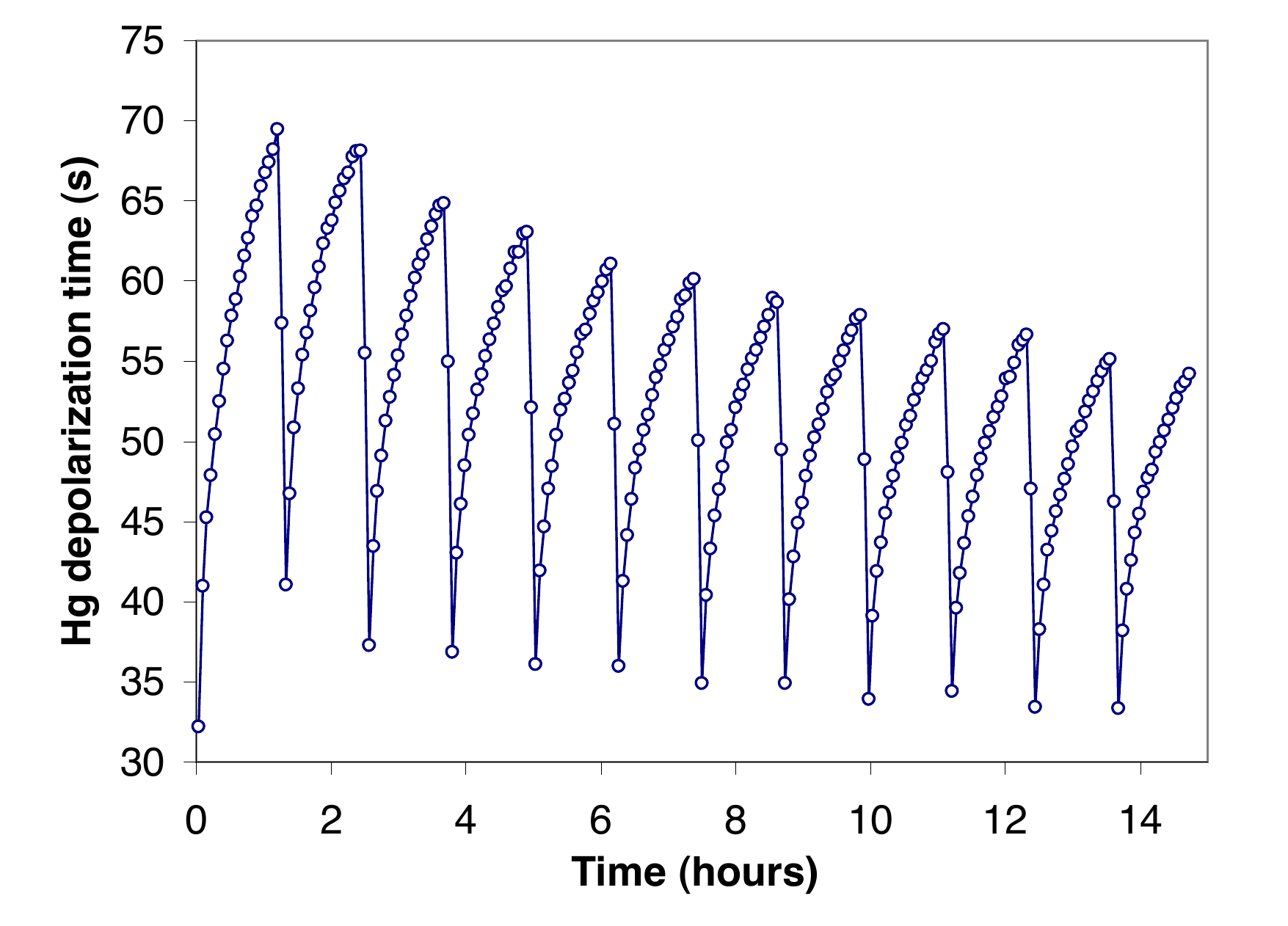}}
\end{center}
\caption{(Color online) Behavior of the mercury depolarisation time
during repeated reversal of the applied electric field.}
\label{fig:Hg_HV}
\end{figure}

The depolarization time of the mercury
depended strongly upon the high voltage behavior of the storage cell. As the
upper electrode was charged up, the mercury depolarization time dropped
precipitously, after which it slowly recovered over a timescale of about an
hour. Discharging and recharging at the same polarity had little effect, but
charging at the opposite polarity once again shortened the depolarization
time. During a normal EDM run, the polarity was reversed about once per hour.
The depolarization times therefore followed a characteristic pattern of a series of rapid
falls followed by slow recoveries, upon which was superimposed a gradual
overall reduction, as shown in \figabbr~\ref{fig:Hg_HV}. Sparks also
caused a rapid depolarization, from which there was only partial recovery.

This effect of a temporary increase in
 relaxation each time the HV polarity is reversed may be due to
protons (H$^+$ ions) appearing on the newly positive electrode. Electron 
migration in the dielectric surface layer soon takes over, and the protons
diffuse back into the surface layers again with a characteristic $\sqrt{t}$ dependence as
the HT dwell progresses.  Protons are believed to catalyse the mercury depolarisation 
by forming the paramagnetic short-lived (10$^{-6}$ s) HgH molecules in surface 
encounters.

The
depolarization time could be restored to a large extent by a high-voltage
discharge in 1 torr of oxygen; it was normally necessary to carry out this
procedure every 1-3 days. Prior to this cleaning, the system was usually
``trained'' by increasing the voltage to a fairly high value (between 120
and 170 kV) and allowing it to settle until it could stay for several minutes
without discharging, as discussed in \secabbr \ref{sec: electric field}
below. Cleaning the quartz ring and then heating it in 10$^{-2}$ torr of He at
60 $^{\circ }$C for about two days was also beneficial to the depolarization
time; this procedure was carried out between reactor cycles.

This detremental effect of the high voltage upon the mercury depolarization time could result in a false EDM signal if (a) the average depolarization time were different for the two HV polarities, and (b) the mercury frequency had some small dependence upon the depolarization time.  The cycle-by-cycle dependence of the neutron-to-mercury frequency ratio $R$ upon the mercury depolarization time $\tau$ was found to be $\partial R/\partial \tau =(-0.5\pm3.2)\times10^{-4}$ ppm per second for negative HV and $(+2.2\pm3.3) \times10^{-4}$ ppm per second for positive HV, with an overall average of $(+0.9\pm2.3) \times10^{-4}$ ppm per second.  Coupled with a difference in average depolarization times (where the average has been calculated by weighting with the EDM measurement uncertainties) of $\tau_- - \tau_+ = 2.0\pm 0.2$~s, an average effective neutron frequency shift above 12 nHz may be excluded at 90\% CL.  Such a frequency shift might conceivably contribute a false EDM signal of up to $1.2\times 10^{-27}$ \ecm.  However,  this effect will cancel upon reversal of the magnetic field.  As the quantities of data (as measured by the statistical uncertainty) for the two field directions were identical to within 1\%, an error from this source is excluded at the level of $1.2\times 10^{-29}$ \ecm.

This behavior of the mercury strongly constrained the sensitivity of the
experiment, as it limited the magnitude of the electric field to a value substantially 
below the limit that could be set by leakage currents alone.

\subsubsection{Mercury light shift}
\label{sec:lightshift}

The presence of the Hg reading light, via the Ramsey-Bloch-Siegert mechanism \cite{ramsey55,bloch40}, shifts the resonant frequency of the Hg atoms.  These so-called light shifts are produced  \cite{cohen-tannoudji72, corney}  by any small component, parallel to $\vecB$, of the $^{204}$Hg probe light beam passing through the precessing $^{199}$Hg atoms. This component, and the consequent shift in the neutron-to-mercury frequency ratio $R$, reverse sign on reversal of $\vecB$. A slight dependence of $R$ on the incident light intensity was indeed observed in this apparatus, the magnitude $\sim0.2$ ppm being in agreement with theory. Any changes in intensity correlated with the electric field direction would then result in a frequency shift that would mimic an EDM.  This is the so-called direct light shift discussed in \cite{baker06}.  It is possible to modify the optics to reduce the amount of light travelling in the direction parallel to $B_0$, and in fact this has recently been carried out by the current users of this apparatus.

The method of establishing the light-intensity dependence was as follows.  The reading-light beam can contain several different wavelength components, only one of which serves to measure the $^{199}$Hg precession.  The raw intensity $I_0$ of the light cannot therefore be used to measure any effect of intensity upon $R$.  Instead, the amplitude $a$ of the AC component of the light was used; but it was necessary first to correct it for absorption.  The amplitude as a function of absorption has a characteristic shape that peaks at approximately 16\% absorption over a wide range of light intensities.  It was therefore possible to adjust the measured $a$ for each batch cycle within a run in order to bring it back to its peak value, i.e.\ the value $\a16$ that it would have at 16\% absorption.  

The signal amplitude as a function of the absorption $A$ and polarization $P$ is approximately \cite{green98} (c.f. \eqnabbr \ref{eqn:absorption_defn})
\begin{equation}a = I_0\left(1-A\right)\left\{\left(1-A\right)^{-P}-1\right\}.\end{equation}
This analysis was restricted to polarizations between 5\% and 40\%, and to absorptions greater than 5\%, for which this parameterization is appropriate.

The polarization achieved depends upon the quantity of $^{199}$Hg within the trap, due to the relaxing effect of reemitted photons: the probability of absorbing a reemitted photon increases linearly with the Hg density.  Secondary (and higher) reabsorptions increase even more quickly.  In consequence, $P$ has an exponential dependence upon $A$: 
\begin{equation}P = P_0 e^{-A/A_0}.\end{equation}
Fitting this function within each run yields a characteristic value $P_{16}$ for the polarisation that corresponds to 16\% absorption.  The absorption-corrected amplitude $\a16$ for any $A$ and $P$ within that run is then given by 
\begin{equation}\a16 = a\frac{0.84\left(0.84^{-P_{16}}-1\right)}{(1-A)\left\{(1-A)^{-P}-1\right\}}.\end{equation}

A linear dependence of $R$ on $\a16$ was fitted for each run.  Over the six-year period of data taking for which this apparatus was used, two separate neutron traps were used: one had a rough wall, the other smooth.  A weighted average of the resulting slopes was calculated for each of these two traps and for each direction of $\vecB$.  

An effect of this kind, if present, is expected to be of the order of a fraction of a part per million and is expected to change sign upon reversal of $\vecB$.  The consistency of observed results suggested that it was appropriate to average the results from the two field directions, to obtain a magnitude of $0.21\pm0.08$ ppm/V for the rough trap.  The results for the smooth trap, $0.01\pm 0.03$ ppm/V, were consistent with zero.  It was then possible to correct the rough-trap data for this effect on a run-by-run basis, using the average $\a16$ for the run in question.

The amplitude $a_{16}$ was also observed to have a slight dependence upon the applied HV, as follows:
\begin{itemize}
 \item{}	For $\vecB$ up: $\partial \a16/\partial V = 4.7 \pm 1.2 \times10^{-6}$ volts per kV of applied HV
	\item{} For $\vecB$ down: $\partial \a16/\partial V = 11.0 \pm 1.6 \times10^{-6}$ volts per kV of applied HV
	\item{} Average: $\partial \a16/\partial V = 8 \pm 1 \times10^{-6}$ volts per kV.
\end{itemize}

However, since  it is possible to correct this dependence to within its uncertainty, no net bias should be expected from this source.  There remains an uncertainty on the dependence of $R$ on the HV of $(\partial \a16/\partial V)\times(\partial R/\partial\a16)$ = $3\times10^{-7}$ ppm per applied kV when averaged over both data-taking traps.  The light-signal amplitude dependence on HV is independent of the sign of $\vecB$, but the light-induced frequency shift does change sign with B, and this effect will therefore not cancel upon reversal of $\vecB$.  
For a trap of height $H = 12$ cm this effect therefore contributes an uncertainty of 
\begin{equation} \frac{h}{2}\frac{\partial \nu}{\partial E} = \frac{h\nu}{2}\frac{\partial R}{\partial V}H = 2 \times 10^{-28}\, e\,{\rm cm}.\end{equation}

\subsubsection{Accuracy of Hg frequency measurements}
\label{sec:Hg_accuracy}

A number of mechanisms can affect the frequency measurement of the Hg magnetometer.  Although these do not necessarily have a direct impact upon the EDM measurement, we summarize them here for completeness.

First, an analog of the Bloch-Siegert-Ramsey shift is the light shift due to virtual transitions 
caused by the probe light beam. The size of this effect is estimated to be 0.15 ppm, both by calculations from  
first principles and as assessed in the data by looking for 
frequency versus light intensity correlations: the latter analysis was used, as described above, to correct for this shift.  

Next, there is a real transition shift caused by the fact that
about 10 \% of the Hg atoms used to measure the final phase have been excited 
once before. In the excited state they precess backwards through about 1$^\circ$,
and some of the polarization survives the excitation and decay. The effect is as 
though the gyromagnetic factor and precession frequency were reduced by 0.1 ppm in the auxiliary 
trap and by 0.25 ppm in the data-taking trap. These shifts are expected to be 
completely unchanged by the reversal of $B_0$. 

The total Hg absorption of the light beam is typically 15\%, which gives us
a nearly optimum signal-to-noise ratio. Each atom that absorbs a photon is 
depolarized after the subsequent spontaneous decay ($\tau = 1.2\times10^{-7}$ s). The 
ensemble spin depolarization rate from this cause is about 1/1800 s. The typical 
observed total spin depolarization rate is 1/60 s. The contribution from magnetic-field 
inhomogeneity is expected to be about a hundred times less than that of the 
neutrons (1/600s) making it a negligible 1/60000 s. The  
dominant relaxation rate, close to 1/60 s, is due to spin relaxation when the 
Hg atoms stick on the wall.

The Hg initial phase is established by the two-second 90$^\circ$ spin-flip using a rotating 
field at 8 Hz. Each Hg atom makes about 2000 free paths in the trap during the 
spin-flip, so the phase information is very uniformly implanted over the 
trap volume. It continues to become more and more uniformly spread by 
the Hg motion while the neutrons are flipped using rf at 30 Hz. The initial Hg phase 
is then sampled on the basis of the 1\% of Hg atoms that absorb a photon from the 
light beam during the next 15 s. (These atoms are partly depolarized in the process.)
The final Hg phase is determined from the Hg atoms that absorb a 
photon in the last 15 seconds before the second UCN spin flip. There are  
a number of reasons why the Hg frequency, thus determined, does not represent a perfect
volume average of the field:

\begin{enumerate}
\item{} Finite volume of the light beam: For all the Hg atoms that absorb a 
photon to measure the final phase, the last 1 millisecond of trajectory must 
certainly be near the light beam. This creates a phase bias. The $B_0$ field
near the light could be different by 10$^{-3}$ fractional compared with the 
volume average. This is 0.01 ppm of the total phase previously accumulated. 
The bias should be the same both for the intitial and final phase measurements,
so that it cancels out. The overall shift is expected to be less than 0.001 ppm.

\item{} Artefacts: The system of determining the frequency in the light detector
signal has been tested at the 0.1 ppm level by feeding in sine waves from the 
frequency synthesiser.

\item{} Bias from Hg atoms dwellng on the wall:  Free path transits take about 
10$^{-3}$ s. The sticking time on the wall is thought to be about 10$^{-8}$ s. Thus, the 
overall average has a surface average weighting of 10$^{-5}$ compared to the volume 
average. The surface average value of $B_z$ may differ by one part per thousand from the volume average, 
causing an overall error of about 0.01 ppm.
 
\item{} Bias due to surface relaxation or differential loss: This may occur if the relaxation is faster on 
one wall than another, or if there is a loss of atoms preferentially at one end of the cell. Suppose, for example, that the roof has an excess relaxation rate of 1/100 s 
compared with the other surfaces. Each atom is colliding at about 1000 Hz, 
of which 250 Hz is on the roof. The probability of depolarization per roof 
collision is thus $P = 4\times10^{-5}$. We have analysed this problem and find that a 
shift occurs in the centre of measurement. Under the most pessimistic 
assumptions the shift $\Delta h$ can reach the value $(H/8)P$, where $H$ is the trap 
height -- in this case, $(5\times10^{-6}) H$, or $6\times10^{-4}$ mm.   When the magnetic field is not trimmed, the maximum $\partial B_z/\partial z$ gradients are $10^{-5}$ fractional per mm,  or 1 nT/10 cm. The 
systematic bias from $\Delta h$ is thus 0.006 ppm.

\item{} False EDM due to surface relaxation: The temporary increase in
 relaxation observed each time the HV polarity is reversed has been discussed above.
 This process swings some of the depolarisation rate backwards and 
forwards from roof to floor in synchronism with the HV polarity change. 
This can create a false EDM signal via a finite $\partial B_z/\partial z$. The transient partial 
relaxation rate averaged over an HT dwell is observed to be about 1/100 s, making a 
displacement of $6 \times 10^{-4}$ mm. In the case of a $\partial B_z/\partial z$ gradient of 0.35 ppm/mm
(corresponding to an $R_a$ shift of 1 ppm), the systematic false field 
change seen by the Hg magnetometer is about $2\times10^{-4}$ ppm or $2\times10^{-16}$ T.
This corresponds to a false EDM of about $2\times 10^{-27}$ \ecm, some 1/20th of the 
geometric-phase false EDM. In practice it would have the same signature 
as the geometric phase false EDM, being proportional to $R_a$ and changing 
sign with the direction of $B_0$. It would simply act to increase the gradient of 
both data lines by about 5\%. Currently the lines have a fitted gradient that is
 20\% +=15\% above the GP phase theoretical prediction. This additional effect 
could easily be present. All of its consequences have been covered by our GP 
corrections.

\item{} Finally, variation of light intensity with HV has been dealt with above.  If there were preferential depolarization of Hg on, say, the positive electrode, thus biasing the volume-averaged frequency measurement, it could slightly alter the gradients of the lines in Fig. 2 of \cite{baker06}, similarly to other gradient-changing mechanisms listed; but again, it is not a cause for concern as it does not affect the outcome of the analysis. 
\end{enumerate}

\subsection{The magnetic field}

To carry out a magnetic resonance experiment one must impose conditions on
both the homogeneity and the time stability of the magnetic field: the field
must be sufficiently homogeneous to retain polarization of the neutrons
until the end of the storage time, and it should be sufficiently stable so
as not to increase significantly the uncertainty in the determination of the
precession frequency beyond that due to neutron counting statistics.

\subsubsection{The magnetic shield}

In the environment of the experimental area magnetic field changes of up to $1~\mu $T in a few tens of seconds are quite common, and are often associated
with movements of the reactor crane, or with the operation of magnetic
spectrometers. To provide the required homogeneity and stability of the
magnetic field, the neutron storage volume was set inside a four-layer
$\mu$-metal magnetic shield. The dimensions of the magnetic shield layers are
given in Table~\tablshielddimensions. The two inner layers and their
detachable endcaps had welded joints, and were annealed in a reducing
hydrazine (N$_2$H$_4$) atmosphere at 1050 $^{\circ }\mathrm{C}$ after
manufacture \cite{mag_shields}. The two outer layers, which were too
large to have been fired in a single piece, were made from sheets of
$\mu$-metal individually annealed and bolted together with 150~mm overlaps.
All four layers had  210~mm diameter holes
at the top and bottom of the mid-plane of the central cylinder: The bottom hole contained the neutron
guide tube, and the top contained the high-voltage feedthrough. The endcaps of the
innermost layer had a 45~mm hole in the center, and each of the other three
layers had a 32~mm hole. Originally, the apparatus had been built with a fifth,
innermost, layer of shielding, which was removed in the meantime to allow for
enlargement of the storage vessel. The shielding factor for the set of five
shields was measured by winding a pair of coils around the external shield
frame and measuring the magnetic field change at the center of the shields
with three rubidium magnetometers. The dynamic shielding factor to external
magnetic field changes was found to be approximately ${2\times 10^5}$
radially and ${2\times 10^4}$ axially \cite{sumner87}. With the
four-layer shield, the shielding factor transverse to the axis is approximately $1.5\times10^4$, consistent with expectation and also with comparisons made between changing external fields and changes
registered by the mercury magnetometer.

\begin{table}[tbp] \centering%
\begin{tabular}{|l|l|l|l|l|l|}
\hline
Shield & $R$ (m) & $l_{1}$ (m) & $l_{2}$ (m) & Overlap (m) & $t$ (mm) \\ 
\hline
1 & 0.97 & 2.74 & 2.74 & 0.20 & 1.5 \\ \hline
2 & 0.79 & 2.30 & 2.30 & 0.20 & 1.5 \\ \hline
3 & 0.68 & 0.75 & 1.89 & 0.12 & 2.0 \\ \hline
4 & 0.58 & 0.75 & 1.63 & 0.12 & 2.0 \\ \hline
\end{tabular}
\caption{The dimensions of the four-layer magnetic shield.  Each layer consisted 
of a central cylinder, of radius $R$ and length $l_1$, and two detachable endcaps.
The length $l_2$ is that of the central cylinder plus the endcaps, when assembled.
The overlap is the distance by which the endcaps overlapped the central cylinders.  $t$ is the 
thickness of the $\mu$-metal used in both the central cylinders and the endcaps.}
\label{Tbl: shield sizes}
\end{table}

\subsubsection{The magnetic field coil}

The coil to generate the 1~$\mu $T static magnetic field $\vecB$ was
wound and glued directly onto the aluminum vacuum vessel. The coil fitted snugly inside
the innermost layer of the magnetic shield and was wound with a $\cos \theta $
distribution to give a constant number of turns per unit distance
along the vertical diameter of the cylinder. Theoretically a coil of
constant pitch wound on the surface of a cavity inside a material of
infinite permeability produces a homogeneous magnetic field, regardless of
variation in the cross-sectional area of the cavity. The coil winding used
here was an approximation to this ideal state. The turns were wound 20~mm
apart, and access to the neutron trap required breaking all of the turns
in order to remove the end of the cylinder. Every turn on the coil,
therefore, had two breaks on each end face of the cylinder, where the
electrical connection was made with a brass screw and two brass solder tags.
The magnetic field was aligned with the vertical diameter of the cylindrical
shield, rather than along the axis, to take advantage of the fact that the
radial magnetic shielding factor is greater than the axial shielding factor.

The choice for the magnitude of $\vecB$ was arbitrary, in the sense
that it does not enter directly into the expression for the sensitivity of
the experiment. There are, however, a number of other factors that had a
strong bearing on the choice, \textit{viz}: the field should be large
compared to any residual fields inside the trap ($\leq 2$ nT), so that the axis of
quantization for the neutrons, which is determined by $\vecB$, is in
the same direction everywhere; the field should be large enough to prevent
depolarization of the neutrons as they pass into the shields; the
homogeneity requirements given below must be fulfilled (in general, field
gradients increase linearly with the field itself, thus placing a limit on
the maximum field); the field should be as stable as possible, which is
generally easier to achieve at lower fields; and finally, it was desirable to
keep the precession frequency away from the 50~Hz mains frequency. The 1~$
\mu $T magnetic field chosen in this case gave a resonant frequency of
about 30~Hz for the neutrons.

The coil that generated the $\vecB$ field had a resistance of
approximately 10 $\Omega $, and required a current of 17 mA to provide the 1 
$\mu $T field. The stabilizer providing this current contained a precision voltage 
reference with a very low output voltage temperature coefficient 
(National Semiconductors LM169B; 1 ppm/$^{\circ }$C) and an operational amplifier with a very 
low input offset voltage temperature coefficient (Analog Devices OP177A; 0.03$\mu$V/$^{\circ }$C).
High-stability precision wire-wound resistors (3 ppm/$^{\circ }$C) were used
to define the $\vecB$ field current. High thermal conductivity resin was
used to connect the components to the inside of a cylindrical aluminum block
(approximately 100 mm in diameter and 100 mm long). This block, which acted
as a heat reservoir for the temperature-critical components, was thermally
isolated from the surroundings and from the power supply by more than 100 mm
of polystyrene foam. The average electrical potential of the coil was
maintained at the same potential as the vacuum tank upon which it was wound, 
in order to minimize currents to the coil supports.

\subsubsection{Homogeneity}

\label{section: B0 homogeneity}The homogeneity requirement for a magnetic
resonance experiment in a low-field region is given by Ramsey 
\cite{ramsey84}, following the theory of the hydrogen maser 
\cite{kleppner62}. Consider the neutron storage volume to be characterized by a
length $l$ and to consist of two regions of magnetic field that differ by $
\Delta B$. If $\gamma=-2\mu _n/\hbar $ is the gyromagnetic ratio, then neutrons with velocity $v$ passing from one field region
to the other experience a relative phase shift of 
\begin{equation}
\delta \phi =\gamma \Delta Bl/v.
\end{equation}
In a storage time $T_s$, the neutron will experience $M=vT_s/l$ such phase
shifts, which will add randomly, so that the phase spread during the storage
time is 
\begin{equation}
\Delta \phi \approx \delta \phi \sqrt{M}=\gamma \Delta B\sqrt{lT_s/v}.
\end{equation}
At the end of the storage time $\Delta \phi $ represents the typical phase
difference between any two neutrons arising from them having followed
different paths across the trap. Maintaining polarization requires that $%
\Delta \phi <1$, from which arises the homogeneity constraint 
\begin{equation}
\Delta B_0<{\frac 1\gamma }\sqrt{\frac v{lT_s}}.
\end{equation}
It should be noted that it is the absolute inhomogeneity of the field $%
\Delta B_0$ that is constrained, and not the relative homogeneity $\Delta
B_0/B_0$. Taking $v=5~\mathrm{ms^{-1}}$, $l=150$~mm, $T_s=150$~s and $\gamma
=1.8\times 10^8~\mathrm{radians\,s^{-1}\,T^{-1}}$, the limit becomes $\Delta
B_0<3$~nT. For a $B_0$ field of 1~$\mu $T this requires a relative
homogeneity of $\Delta B_0/B_0<3\times 10^{-3}$ over the 20-liter neutron
storage volume.

The magnetic field within the storage volume was mapped using a
three-axis fluxgate magnetometer probe \cite{bartington}. Spatial variations
were found to be of the order of the $\approx $1 nT resolution of the
instrument, as long as the shield was demagnetized each time that the
magnetic field configuration changed (i.e., each time the magnetic shield was
opened or the direction of $\vecB$ was reversed). Demagnetization was
carried out by using a current loop that was threaded through all of the
shields, parallel to the cylinder axis. The current was initially set to 100
amp\`{e}re-turns, reversed every 2~s and steadily reduced to zero over
twenty minutes. Trim coils were used to achieve this level of homogeneity;
without them, the field variations would have been about four times greater. 
The $T_{2}$ neutron polarization relaxation time was typically about 600 s; the field inside the
trap was therefore adequately homogeneous to meet the requirements of the
experiment.

\subsubsection{Stability}

\label{sec: mag field stability}In order to ensure that any noise on the EDM
signal caused by magnetic field instabilities was significantly less than
that due to neutron counting statistics, it was required that the shift in 
precession frequency between consecutive measurements should normally 
be not much larger than the uncertainty due to neutron counting statistics.
Thus, 
\begin{equation}
\left| {\frac{dB_{0}}{dt}}\right| {\frac{\gamma T_{s}}{2\pi }}\lessapprox {%
\frac{1}{2\pi \alpha T_{s}\sqrt{N}}}.  \label{eqn: stability condition}
\end{equation}
For $\alpha =0.5$, $T_{s}=130$~s, and $N=10\,000$ the constraint therefore
becomes $|dB_{0}/dt|\lessapprox 8$ fT/s. For $B_{0}=1~\mu 
\mathrm{T}$, this requires a stability of about one part per million over
130~s. However, this criterion is stricter than was necessary in this
instance, for two reasons. First, the separated oscillating field method
itself is relatively insensitive to fluctuations in the magnetic field on
time scales short compared with $T_s$. This is because the
neutron counts are determined by the total accumulated phase difference
between the neutron polarization and the oscillator, and not by a detailed
comparison throughout the storage cycle. Second, the measured mercury
precession frequency was used for normalisation. Except for a period of about
5\% at either end of the storage time, any drifting of the magnetic field
affected both spin systems in exactly the same manner, and averaging
over the entire Ramsey measurement period reduced the influence of any changes that did
occur during the end mismatch periods by an order of magnitude. In practice, though,
condition (\ref{eqn: stability condition}) was usually satisfied. On the rare
occasions when the field changed much more rapidly than this, the mercury
precession was generally disturbed to such an extent that $\chi ^{2}/\nu $
for the frequency fit became extremely large, and the data point was
rejected.

\subsubsection{Uncompensated magnetic field fluctuations}
In principle it is possible to have residual effects from ${\bf B}$ field fluctuations, such as hysterisis in the $\mu$-metal shield following disturbances in the stabilised 
$\vecB$ coil current supply caused by pickup from the high voltage changes.  This would manifest itself most strongly as a dipole-like field ${\bf B}_d$ originating from the $\mu$-metal in the region of the HV feedthrough, which would be sensed by both the neutrons and the mercury magnetometer but with a difference given by $\delta B_d/B_d = 3\Delta h/r$ where $r\sim 55$ cm is the distance from the source of the field to the center of the trap.  Thus, fluctuations in ${\bf B}$ that are correlated with the HV can be expected to be compensated up to a factor of about 70.  In order to study this, the mercury and neutron channels were analysed independently.  

\begin{figure} [htb]
\begin{center}
 \resizebox*{0.5\textwidth}{!}{\includegraphics[clip=true, viewport = 53 420 533 760]
{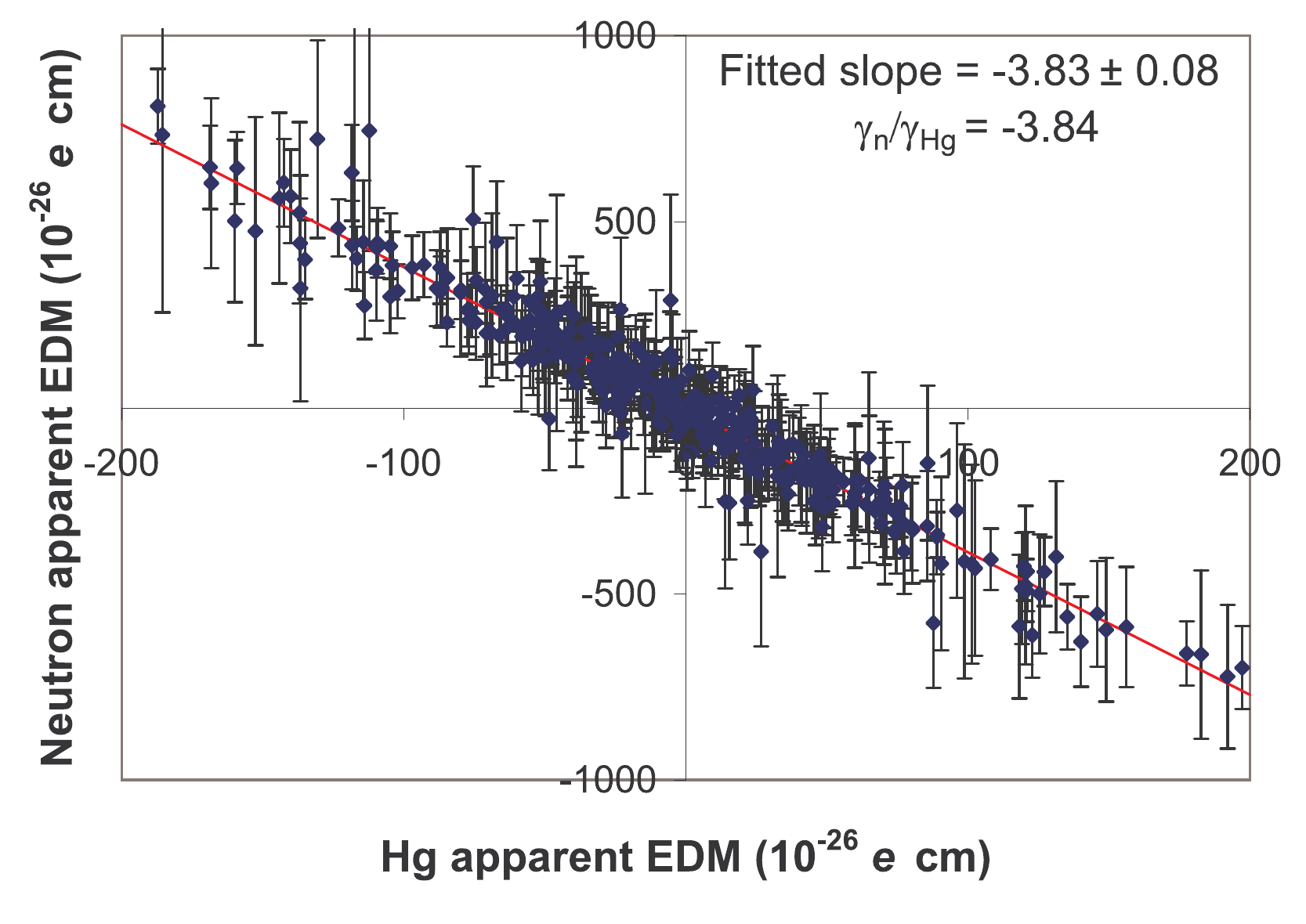}}
\end{center}
\caption{(Color online) Apparent neutron EDM signals (due to
uncompensated random magnetic field fluctuations) as a function of the
corresponding apparent mercury EDM signals.}
\label{Fig:n_vs_Hg}
\end{figure}

The analysis was performed by selecting sequences of measurement cycles within each run for which the magnetic field (as measured by the mercury
frequency) varied smoothly throughout several high-voltage dwell periods. Both the mercury and the neutron frequencies for each such sequence were
fitted to a low-order polynomial. The fits were unweighted, since the displacement from the fitted function was entirely dominated by the magnetic
fluctuations rather than by the uncertainties in the frequency calculation associated with each point. The residuals were then fitted to a linear
function of the applied electric field to yield the apparent EDM measurements.  A plot of the neutron vs.\ the Hg results (\figabbr \ref{Fig:n_vs_Hg}) shows complete (within uncertainties) correlation between the results, with the slope of the best-fit line ($-3.83 \pm 0.08$)  corresponding as expected to the ratio of gyromagnetic ratios. 

The neutrons yielded a net uncompensated EDM signal of $(17\pm4)\times10^{-26}$ \ecm; the Hg (once geometric-phase-induced false EDM contributions\cite{pendlebury04} had been subtracted) yielded $(-3.9\pm0.8)\times10^{-26}$ \ecm. These results are consistent with a common source of magnetic fluctuations correlated with the HV.  We therefore expect the mercury-magnetometer compensation to shield us from this systematic effect to a level of $17\times10^{-26}/70 = 2.4\times10^{-27}$ \ecm.

\subsection{The electric field}

\label{sec: electric field}The main requirements for the electric field were
that it should be as large as possible and aligned with the magnetic field,
but with the constraint that the leakage current through the insulator of the
neutron trap should not generally exceed a few nanoamps. This latter restriction arises  
because the magnetic fields produced by currents circulating around
the trap would induce shifts in the precession frequency that were correlated with
the electric field. Although such frequency shifts would be compensated
to the level of at least 90\% by the mercury magnetometer, any residual effect could
result in a systematic error in the EDM, as discussed below.

Sparks could also in principle generate a systematic effect if they changed the
magnetization of the shields and if they occurred preferentially for one
polarity of the electric field.  However, the mercury magnetometer would
naturally compensate for any such effect, just as with any other shifts in
the magnetic field.

Sparks were also undesirable because, as
discussed in \secabbr (\ref{section: Hg and high voltage}) above, they caused
the mercury atoms to depolarize rapidly. The frequency at which sparks occurred
depended upon the voltage used, the quality of the vacuum, and the
conditioning of the system \cite{alston}. Sparks occurred more
frequently when the experiment was under vacuum ($\approx 10^{-6}$~torr) than
they did when a pressure of $10^{-3}$~torr of either dry nitrogen or helium
was maintained in the system. Helium was found to be more efficient than
nitrogen in quenching sparks.

Before the start of each data run, the electric field was raised as far as
possible (typically 1.5 MVm$^{-1}$), maintained for several~minutes, and
then lowered and applied with the opposite polarity. This was repeated
several times. The effect was to reduce both the quiescent current across the
trap and to suppress almost entirely the occurrence of sparks during normal data taking. It
was then necessary to ``clean'' the trap with a short high-voltage
discharge in 1~torr of O$_{2}$ (with a current of 130~$\mu$A for
approximately two minutes at each polarity, twice) in order to restore the
depolarization time of the mercury to a reasonable value. To some extent the
cleaning reverses the beneficial effects of the training, and so the
cleaning period is kept as brief as possible. The maximum electric field
used for data taking was $1~\mathrm{MV\,m^{-1}}$, since occasional
high-voltage breakdowns tended to occur beyond this limit, resulting in a reduction in the mercury depolarization time. 

As sparks invariably disrupt the mercury frequency measurement, batch cycles that
contain them are excluded from the analysis, so beyond the residual effects
just discussed the sparks themselves cannot contribute to any artificial EDM
signals.

\subsubsection{The high-voltage stack}

\begin{figure} [ht]
\begin{center}
\resizebox{0.5\textwidth}{!}{
\includegraphics{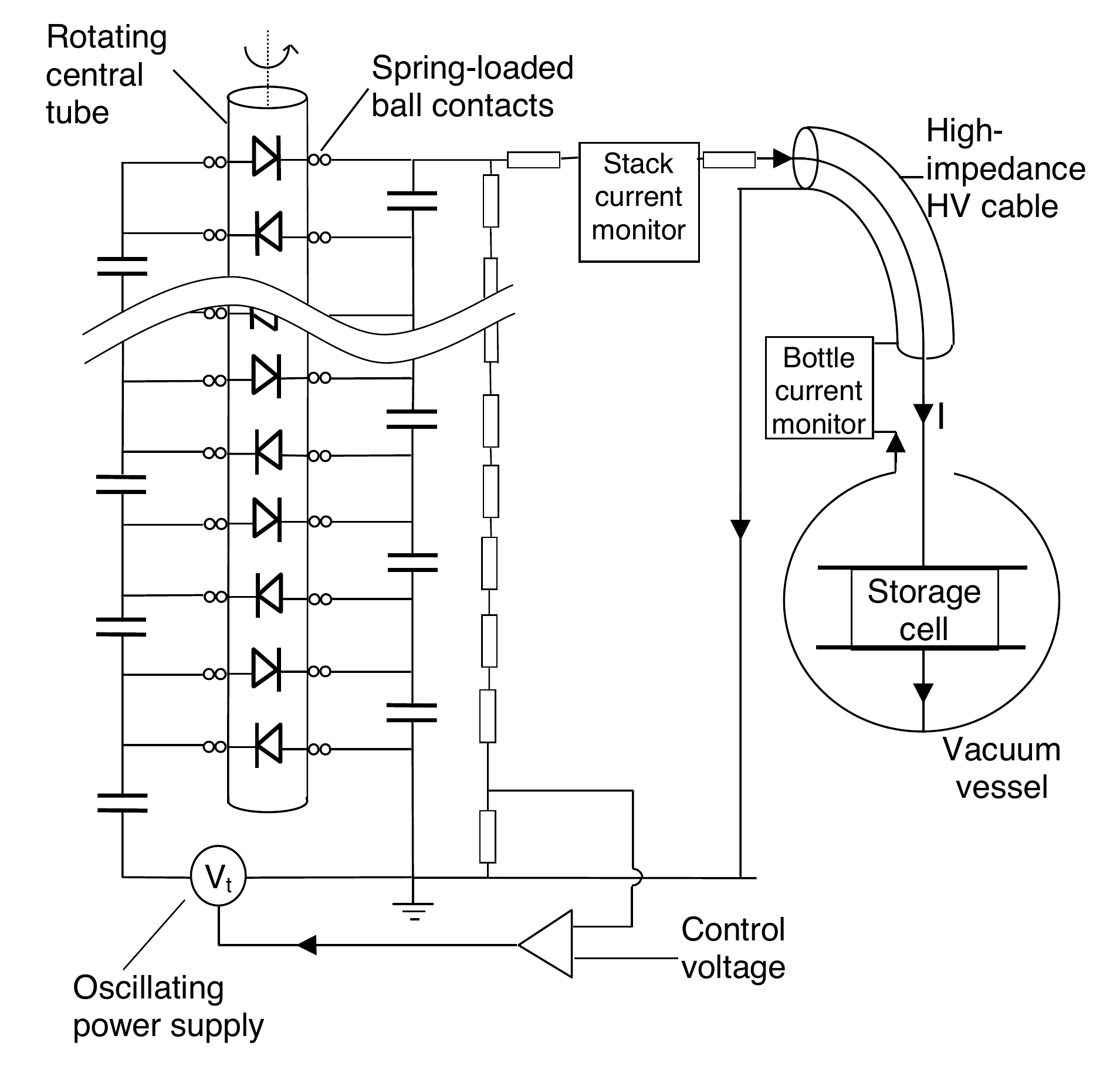}}
\end{center}
\caption{The reversible Cockcroft-Walton type
high-voltage stack, and the current path through the EDM apparatus.}
\label{HV_stack_diagram}
\end{figure}

The electric field was generated by a reversible Cockcroft-Walton type high
voltage stack, shown schematically in \figabbr~\ref{HV_stack_diagram}.
The stack was powered by a controller from Bonar Wallis \cite{bonar_wallis_ltd}.

The polarity of the electric field within the neutron trap was reversed by
changing the sign of the voltage applied to the ungrounded electrode. This
was done by physically reversing the diodes in the charging stack, with the
stack at zero voltage. The reversal was driven, under computer control, by a
180$^{\circ }$ rotation of the core of the stack using compressed air. The
stack was connected to the neutron trap by 5~m of coaxial
high-voltage cable with its central conductor removed and replaced with oil.  
A semiconducting sheath around the central conductor remained, and this provided the 
primary conducting path through the cable. 
There was a 1~G$\Omega $ resistance in series between the
cable and the trap, to limit the current.

The stack, which was capable of providing $\pm $300~kV, was driven by a 20~kHz
oscillator connected to the lowest of its 15 stages. Each stage was separated
from its neighbors by a 3.6~nF capacitor, and a return current through a
2.8~G$\Omega $ resistor chain from the top stage was used by the controller
to stabilise the output voltage.

\subsubsection{Monitoring the high voltage}

The electric field in the trap was monitored by recording the magnitude and
sign of the voltage at the top of the stack, the current flowing through the stack, and the current
in the feedthrough just above the trap, which charged up the electrodes (primarily displacement
current) as the electric field was changed. \figabbr~\ref{HV_stack_diagram}
shows schematically how the current through the neutron trap was monitored.
The coaxial arrangement of the trap and the return current path ensured
that the magnetic effects of this current were minimized. This design arose
from the experience gained in the earliest version of this experiment: At
that time, the vacuum vessel was a glass jar, and no coaxial return current
path was available. Sparks within the experimental apparatus were then seen
to magnetize the shields permanently, producing changes of as much as 1~mHz
in the precession frequency of the neutrons. With the arrangement described here no
such effects were seen in this experiment.

\subsubsection{Leakage currents and their effects}
\label{sec:syst_leakage}

\begin{figure} [ht]
\begin{center}
\resizebox{0.5\textwidth}{!}{
\includegraphics{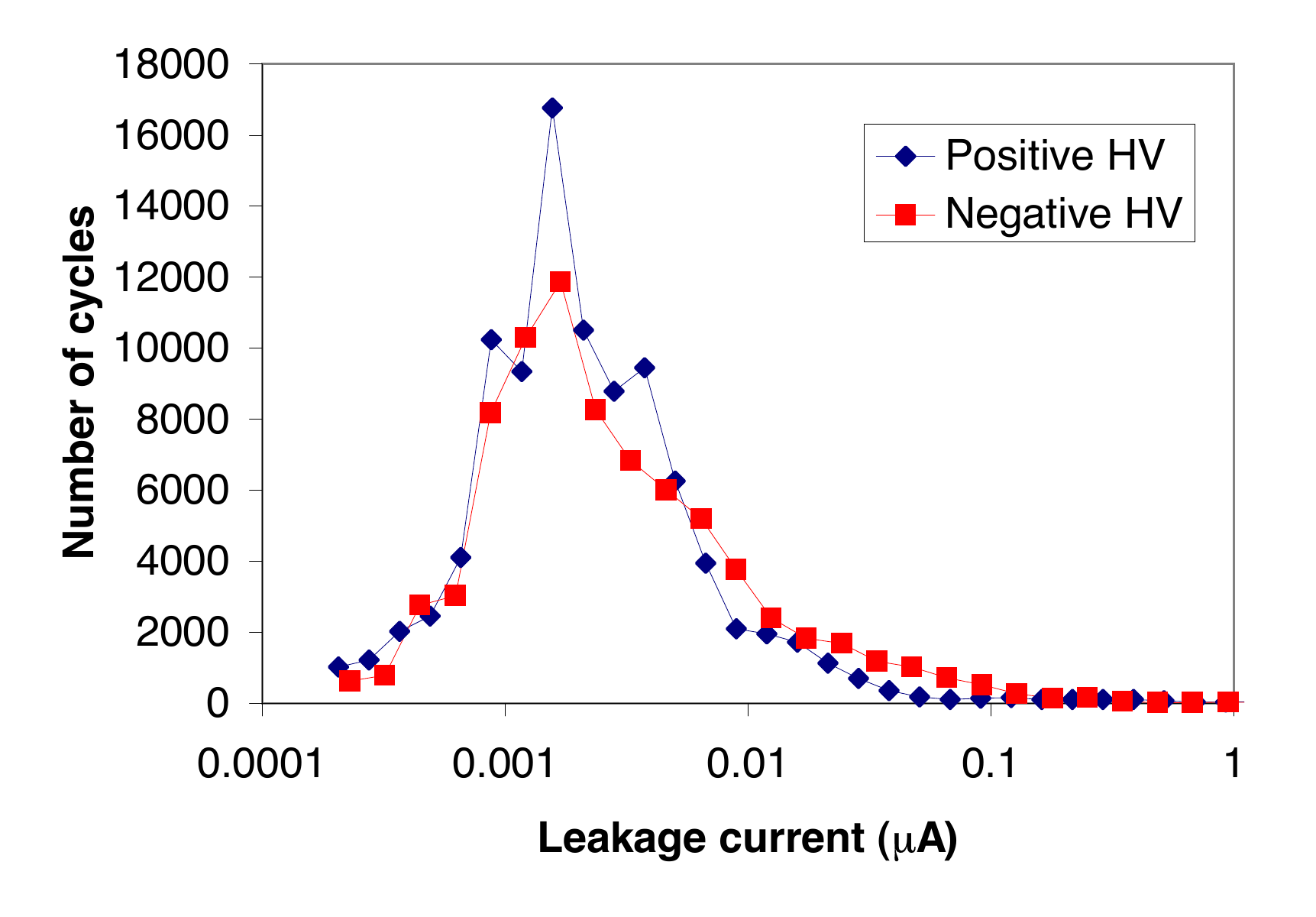}}
\end{center}
\caption{(Color online) Distribution of the  average leakage currents observed during each batch cycle. $I$}
\label{fig:leakage_current_distrib}
\end{figure}

By a suitable choice of the high-voltage setting the quiescent current
through the trap was typically kept at or below a few nA. The distribution for both polarities is shown in \figabbr \ref{fig:leakage_current_distrib}.  

\begin{figure}[ht]
  \begin{center}
    \resizebox*{0.5\textwidth}{!}{\includegraphics
    [clip=true, viewport = 53 436 543 763]
    {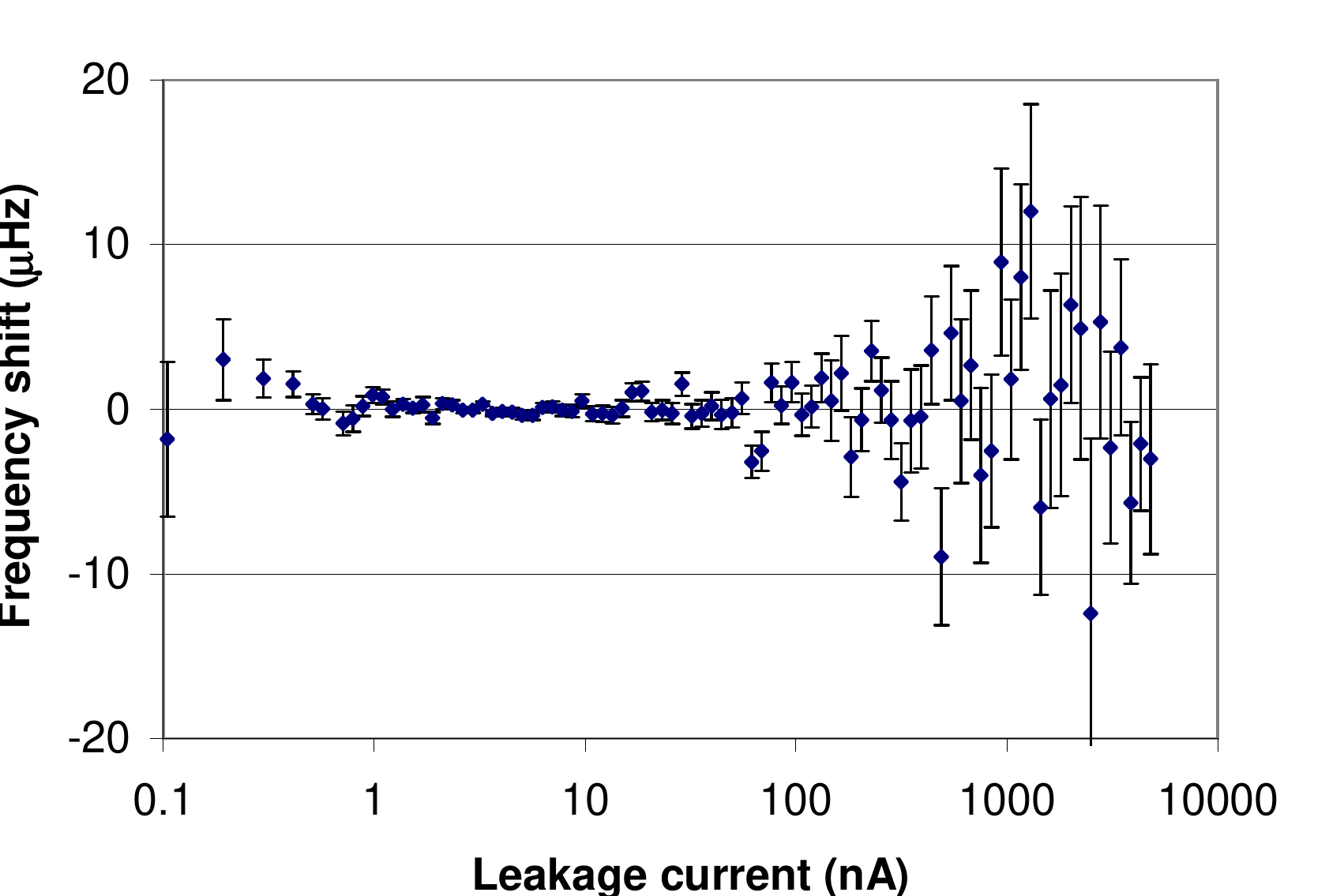}}
    \caption{(Color online)  Frequency shifts (multiplied by the polarities of the electric and magnetic fields) as a function of leakage current.}
    \label{fig:frq_shift_vs_I}
  \end{center}
\end{figure}

If the current flows in an axial direction through (or along the surface of)
the insulator between the electrodes, the magnetic field that it produces
will be at right angles to $\vec{B}_{0}$. This field will be small compared
with $\vec{B}_{0}$ and will produce a shift in the precession frequency that
is independent of the polarity of the electric field; thus, this will not be
a source of error in the measurement of the EDM. However, one cannot assume
that the current will take such a direct path. The insulator is likely to
contain paths of different resistances, which could lead to the current
having a net azimuthal component.  (The insulator ring showed some mild discoloration indicating the path of discharges along its surface.  For the most part these were vertical, but occasionally they were at an angle of up to 45$^\circ$.  It is likely that discharges along the surface of the insulator occurred most often in the vicinity of the windows for the mercury light.) In this case, a component of the magnetic
field due to the current would be parallel (or anti-parallel) to $\vec{B}_{0}$ and
would produce a frequency shift that changes sign when the polarity of the
electric field is reversed, giving rise to a systematic error in the EDM.
This effect can be estimated for the case in which the current $I$
makes a fraction $f$ of a complete turn around the insulator. If the
insulator has radius $r$, the magnetic field at the center of this current
loop is 
\begin{equation}
B=\frac{\mu _{0}I}{2r}\cdot f. 
\end{equation}
The mercury should compensate for the resulting frequency shift at a level
of 90\% or more. The current would therefore generate an artificial EDM
signal of magnitude 
\begin{equation}
|d|=0.1\frac{\mu _{n}}{E}\frac{\mu _{0}I}{2r}\cdot f.
\end{equation}
As shown above, leakage currents are normally of the order of 1 nA.  
If the current travels an azimuthal distance of 10 cm around the 47 cm diameter
trap, the applied electric field of $E = 1$ MV/m would give a false signal of
order $0.1\times 10^{-27}$ \ecm. \figabbr \ref{fig:frq_shift_vs_I} shows the binned weighted-average frequency shifts (i.e., the departures from the fitted Ramsey curves of the individual measurement cycles) as a function of the leakage current.  The frequency shifts are multiplied by the product of the polarities of the electric and magnetic fields.  No dependence on leakage current is apparent.

The displacement current as
the voltage is ramped up and down was typically 1 $\mu $A. The magnitude of
this, along with the known capacitance of the trap, provided the necessary
evidence that the applied voltage was reaching the trap. The current
flowing through the trap was not measured directly. The measured current
included currents flowing in the high-voltage feedthrough and cable
assembly, and it therefore should be regarded as an upper limit for the
current that flowed through the trap.

\subsection{HV AC ripple}

Changes in precession frequency may be caused by oscillating magnetic fields at non-resonant frequencies through Bloch-Siegert-Ramsey type effects \cite{ramsey55}. An example in this class is a ``ripple'' on the high voltage, which would generate an oscillating displacement current in the storage chamber and thereby an  oscillating $B$ field. The ripple amplitude may change with the sign of the high voltage, producing slightly different frequency shifts for each of the two high voltage polarities.

Consider the presence of an oscillating field $\vec{B}_{2}\sin \omega _{2}t$ in addition to the static field $\vec{B}_{0}$ and the resonant alternating field $\vec{B}_{1}$. During the storage time $T_s$, when $\vec{B}_{1}$ is off, the magnetic field in the trap is  
\begin{equation}
\vec{B}_{t}=B_{0}\,\hat{k}+B_{2}\sin \omega _{2}t\,\hat{\imath}.
\end{equation}
For $B_{2}\ll B_{0}$ and $\omega _{2}\gg \mathrm{T_{s}^{-1}}$, the time-averaged magnitude of this field is 
\begin{equation}
\langle B_{t}\rangle \approx B_{0}\left( 1+\left( {\frac{B_{2}}{2B_{0}}}\right) ^{2}\right) .
\end{equation}
The precession frequency therefore becomes 
\begin{equation}
\nu _{0}^{\prime }=\nu _{0}\left( 1+\left( {\frac{B_{2}}{2B_{0}}}\right) ^{2}\right) ,
\end{equation}
where $\nu _{0}$ is the frequency in the absence of $\vec{B}_{2}$. 

The most probable source of an AC~magnetic field is the 20~kHz oscillator that drives the high-voltage stack. This current keeps the capacitors charged against the losses due to the monitoring current. If the driving frequency is $\omega _{2}$ and the monitoring current is $I_{s}$, the voltage associated with this current is 
\begin{equation}
\mathcal{E}={\frac{I_{s}}{\omega _{2}C}},
\label{eqn: voltage of stack-driving current}
\end{equation}
where $C$ is the capacitance of the stack. For the fifteen-stage stack with one 3.6~nF capacitor per stage, $\omega _{2}=1.3\times 10^{5}$ rad/s and $I_{s}=100~\mu$A, equation (\ref{eqn: voltage of stack-driving current}) yields $\mathcal{E}=3~\mathrm{V}$.

The capacitance of the trap, as calculated for a pair of parallel plates, is 15~pF, which, at 20~kHz, has an impedance of 0.5~M$\Omega $. This shorts out the DC resistance of the trap. Between the stack and the trap is a 1~G$\Omega $ resistor chain, so that 3~V produces a 3~nA alternating~current. This current flows through the trap as a displacement current and produces an AC~magnetic field whose magnitude, averaged over the volume of the trap, is of the order of 1 fT. This would give a frequency shift of $\approx $ 10$^{-17}$ Hz and a systematic error in the EDM at the level of $\approx 10^{-36}$ \ecm, which is a completely negligible effect.

AC fields at mains frequency are another possible cause of concern.  There is no differential ripple visible on the HV at the level of a few volts.  Sampling is done at 5 Hz with a bandwidth of 20 kHz, so any 50 Hz ripple would show up as beats.  This is certainly absent at the level of, say, 50 V, which would give a false EDM of $0.01\times10^{-27}$ \ecm.

Low-frequency AC fields were sought by means of a pickup coil in conjunction with a phase-sensitive detector.  Shifts in $R$ from this source at the level of 0.02 ppm could not be ruled out. Cancellations in the corresponding EDM signal from reversals of the electric and magnetic fields would reduce any net contribution to below the level of $0.01\times10^{-27}$ \ecm.

\subsubsection{Electric forces}
Another possible source of systematic error arises from electrostatic forces, which may move the electrodes slightly.  In conjunction with a magnetic field gradient, an HV-dependent shift in the ratio would then appear.  This was sought by looking for an EDM-like signal but with a frequency shift proportional to $|{\bf E}|$ instead of to ${\bf E}$.  The $|{\bf E}|$ signal, at $(-2.4\pm 3.8) \times10^{-26}$ \ecm, was consistent with zero. If the HV magnitudes were slightly different for the two signs of ${\bf E}$, this effect would generate a false EDM signal.  Study of the measured HV and of the charging currents show that the HV magnitude was the same for both polarities to within an uncertainty of about 1\%.  This systematic uncertainty is therefore 1\% of the $|{\bf E}|$ uncertainty, i.e.\ $0.4\times10^{-27}$ \ecm.

\section{The data-acquisition process}

A data-taking run lasted for up two days and involved a sequence of
operations built around the continuous repetition of the basic Ramsey measurement
cycle outlined in \secabbr \ref{sec:ramsey_method}. This cycle lasted for approximately four minutes, and involved filling
the trap with polarized neutrons and mercury, applying the separated
oscillating fields sequence, releasing and counting the neutrons in the
original spin state, and finally releasing and counting the neutrons in the
other spin state. Each cycle gave rise to a single neutron frequency measurement.
Approximately once per hour, the direction of the electric field was
reversed. The operation was controlled by a PC\ running LabVIEW-based
software.\cite{nat_instr}

\subsection{Filling}

The trap was filled for 20~s, corresponding to approximately 1.3 filling
time constants, after which the density of UCN was about $2~\mathrm{cm^{-3}}$. The polarization at this time was approximately 75\%. The stored neutrons
had their spins aligned antiparallel to the magnetic field in the trap
(denominated  ``spin up''). At this point the neutron door was closed, and the door
from the mercury prepolarizing cell was opened for 1~s, allowing the
polarized mercury atoms to enter.

\subsection{Ramsey sequence}

The Ramsey sequence then began, with a 2~s interval of rotating magnetic
field ${\bf B}_{1}^{\prime }$ (in the horizontal, or $xy$, plane) to allow the 
mercury polarization to precess down into the $xy$
plane, followed by a 2~s interval of (horizontal) oscillating field $\vec{B}_{1}$ to turn
the neutron polarization in similar fashion. The ${\bf B}_{1}$ field was
aligned with the cylinder axis of the shield and it was generated by a
Helmholz pair of current-carrying wire turns on the vacuum vessel. The
current was provided by an HP 3325B frequency synthesiser\cite{hp}. The
inner magnetic shield acted as a return for the flux. The ${\bf B}
_{1}^{\prime }$ field (for the mercury) was a superposition of two perpendicular linear
oscillating fields, 90$^{\circ }$ out of phase, generated in an identical
manner by their own Helmholtz pairs. The simple nature of the coils, and the
distorting effects of eddy currents in the vacuum chamber wall and other
metal parts, caused the oscillating field to vary in strength by about 10\%
over the volume of the neutron trap. Conveniently, the rapid motion of the
mercury and neutrons inside the trap provided sufficient averaging in the
2~s duration chosen for each r.f.\ pulse interval that, in spite of this
inhomogeneity, there was a negligible loss of polarization while turning the
polarization vectors into the $xy$ plane. The fact that the neutrons remained
relatively undisturbed during the four-second period after the closing of
the neutron door and before the ${\bf B}_{1}$ pulse was applied allowed the
neutron velocity distribution to relax towards isotropy, and the spatial
distribution to relax towards uniformity. This should have minimized any
systematic $\vxE$ effect arising from the Lorentz transformation of the electric field
into the neutrons' rest frame.

A 130~s interval $T_{fp}$ followed in which the spin polarizations precessed freely
in the $xy$ plane about the $\vecB$ and $\vecE$ fields. The choice
of the length  of $T_{fp}$ depended upon several factors: (i) the
storage lifetime of neutrons in the trap (about 200 s); (ii) the $T_{2}$
relaxation time of the neutrons (about 600~s, although times as long as 1000
s were seen under the best conditions); (iii) the resulting width of the resonance; (iv) the dead time spent in filling and emptying the trap, since the sensitive period $T_{fp}$ should be as long as possible in comparison to them; (v)
the signal-to-noise and the depolarization time of the mercury, which affect
the accuracy of the frequency measurement; and (vi) the needs of other users
of the TGV neutron source, whose measurement cycles had to be interleaved with
those of the EDM experiment.  The maximum statistical sensitivity was achieved by  maximizing, as far as possible,  the quantity 
$\alpha ET_{fp}\sqrt{N_b/T_{\mathrm{tot}}}$, where $T_{\mathrm{tot}}$ is
the total time taken for the measurement cycle and $N_b$ is the number of neutrons per batch cycle. This function is, in fact, rather flat in the region of the 130 s
storage time that was used.

\subsection{Counting}

The free precession was brought to an end when the frequency synthesiser was
gated on to the coil to provide the second 2~s interval of the oscillating $%
B_1$ field. Immediately afterwards, the door of the trap was opened. The
polarizing foil then served as an analyzer and let through to the detector
only those neutrons that project into their original spin-up state. After 8
s of counting, a fast-adiabatic-passage spin-flip coil, adjacent to the
polarizer, was energized. The spin-down neutrons, which had until this time
been unable to pass the polarizer, then received a $180^{\circ }$ spin flip
whenever they traversed the spin-flip coil. This permitted them to pass through
the polarizer and on to the detector. They were counted in a separate scaler 
for 20~s, before the system reverted to continued counting of the
spin-up neutrons for a final 12~s. Counting the spin-down neutrons served a
triple purpose: it increased the sensitivity of the experiment by increasing
the number of neutrons counted; it emptied the trap of neutrons that
would be in the ``unwanted'' spin state when refilling at the beginning of
the next cycle; and finally, it provided a way of eliminating noise that
would be introduced by fluctuations, additional to those of normal counting
statistics, in the initial number of stored neutrons after filling. The
spin-up and spin-down counts belong to different Ramsey resonance patterns
that are 180$^{\circ }$ out of phase.  Splitting the spin-up counting into two periods and inserting the spin-down counting in between them allowed us approximately to equalise the efficiency of detection of the UCN leaving the trap in each state.

The first batch of any run is different from any of the others, as the neutron trap and guides are initially empty; for other batches there is likely to be some remnant population from the previous batch.  In consequence the first batch often had an anomalously low total neutron count (and would normally be excluded from analysis).

\subsection{Timing}

The timing of the various stages of the measurement cycle was controlled by a
dedicated microprocessor. It was installed as a CAMAC unit so that at the
start of a run the interval lengths to be used, and the corresponding states of the various valves and relays, could be loaded into the
microprocessor memory from the PC that was in overall control of the
data acquisition.   

After it had started a cycle, the PC became completely
passive with respect to timing. It received signals from the timer that told
it the logical state of each hardware control. As each cycle neared its
end, the PC awaited an end-of-sequence signal from the timer, at which
point it immediately restarted the timer sequence for the next cycle. This
ensured that the timing within the cycle, which could potentially influence the number of
neutrons counted, could not be affected by the state of the high voltage in
some unforeseen way through the action of the software. End-of-cycle tasks
such as storing the data on disk and reprogramming the frequency synthesizer
were carried out during the first few seconds of the subsequent cycle.

\subsection{High-voltage control}

The high voltage was controlled by a separate PC, and the associated
controlling and monitoring electronics were kept entirely separate from the
data-acquisition electronics. The PCs were networked via a common Ethernet hub. 
At the start of the run, and after each Ramsey measurement period,
the data acquisition PC issued a request to set the appropriate voltage for the
upcoming batch. The high-voltage PC transmitted in return a summary of
measurements that it had made, such as the average voltage, leakage current,
maximum current and so on, during the Ramsey measurement period that
has just been completed. These data were stored along with all of the other
information relating to that particular measurement cycle. Keeping the high
voltage control separate from the data acquisition system minimised the
possibility of some unforeseen interaction that might result in a false EDM
signal. The initial polarity of the high voltage at the
start of the run was chosen randomly by the software.

The high voltage changed with a pattern that repeated every 32-40 measurement
cycles (collectively known as a ``dwell''), 
the exact sequence being programmed as desired at the start of the
run. There were typically 16 cycles with the electric field applied, say,
parallel to the magnetic field, followed by two or four cycles at zero
electric field, before the sequence was repeated with the electric field
reversed. The electric field did not
normally attain its full value until the second cycle of each dwell, 
because it took a significant amount of time to reverse the
polarity and to ramp up the voltage. Only the 40~s period during which the
neutrons were being counted was used to change the electric field; the voltage
was frozen at the start of the measurement cycle, allowing it to settle
and the leakage currents to fall during the neutron filling period so that
it was stable during the sensitive Ramsey measurement period. Data taken during
the first batch cycle of each high-voltage dwell are therefore valid, and are used in the
analysis, but have a reduced sensitivity relative to the majority of other
cycles because of their lower electric fields. In principle, it would have been
possible to ramp up fast enough to complete the polarity change within one
40~s period, but doing so would have increased both the displacement current and
the probability of sparks occurring.

Thus, the electric field is taken through a cycle of changes that has a
repetition period of about 2 hours. The length of this period was chosen
with the following considerations in view: The magnetic field had slow drift
noise, or what might be called ``$1/f$'' noise, which, if not treated
properly, might have made  a significant contribution to the statistical error on
the measurement of the EDM. The use of the electric field reversal sequence
with a period $T_{E}$ makes it possible to reduce the noise contributions
coming from the spectral components of the drift with period $T_{s}p$ by a
factor which is approximately $T_{s}p$/$T_{E}$. Thus, shortening the period
for the electric field sequence increased the attenuation for the drift
noise at very low frequencies and extended the attenuating effects to higher
frequencies. Furthermore, the system was constrained by the behavior of the
mercury; it was usually necessary to end a run after a day or two in order to
discharge-clean the trap so as to restore the mercury depolarization time. Since it was clearly desirable to have several complete
high-voltage dwell periods within each run, one hour was a reasonable maximum
time limit between polarity reversals. The disadvantage of shorter dwell
sequences is that more time would have been spent at low voltages while the field was
being ramped; and, in addition, the mercury depolarization time took an
hour or so to recover from the dramatic fall that it suffered at each
polarity reversal (see \figabbr~\ref{fig:Hg_HV}).

Study of the measured HV and of the charging currents show that the HV magnitude was the same for both polarities to within an uncertainty of about 1\%. 

\subsection{Neutron frequency tracking}

The mercury frequency $\nu_{Hg}$ for each cycle was used to derive a first-order estimate 
\begin{equation}
\nu _0^{\prime } = \nu_{Hg}\frac{\gamma_n}{\gamma_{\mathrm{Hg}}}
\label{eqn:Hg_to_n_freq}
\end{equation}
 for the neutron resonant frequency.  This allowed the applied synthesizer frequency $\nu_1$ to be adjusted on a cycle-by-cycle basis in order to track variations in the magnetic field.  
The frequency $\nu_1$ was made to differ from $\nu _0^{\prime }$ by an amount
\begin{equation}
\delta \nu =\nu _0^{\prime }-\nu _1
\label{eqn: applied minus expected freq}
\end{equation}
where $\Delta \nu$ is the linewidth given by equation (\ref{eqn: linewidth}) and $f$ was chosen sequentially to be -0.55, +0.45, -0.45, +0.55,  so as to follow the pairs of working points on either side of the central fringe of the resonance as shown in \figabbr~\ref{fig:ramsey_resonance}. 

\subsection{Measurement and storage of data}

The state of the experiment was monitored and recorded using 24-bit scalers
and 12-bit, 10~V ADCs that were read at various points during each
measurement cycle, as well as by the 16-bit ADC used to record the oscillating
mercury signal. The values of about fifty parameters were written to disk for
each cycle. These parameters included the neutron counts for each of the two
spin states; neutron counts registered by the flux monitor on the input
guide tube; the frequency of the applied oscillating $\vec{B}_1$ field; the
fitted mercury frequency, amplitude and depolarization time, with their
associated uncertainties; the high voltage magnitude and polarity; average
and maximum leakage currents during the Ramsey measurement period; and various
supplemental information, such as the temperature and humidity of the
environment. The mercury ADC readings were stored in separate files, in case
the need should arise to reanalyze and refit them. For each run, a
multichannel analyzer (LeCroy\cite{lecroy} qVt module) recorded the
pulse-height spectrum from the neutron detector, and this spectrum was also
recorded on disk so that the performance of the detector could be monitored
over time. In addition, values for the voltage and current in the HV\ system
were digitised at a rate of 5 Hz, and these readings were also recorded
separately so that the high-voltage performance of the system could be
examined in detail for any given run.

A single run typically lasted for one to two days, and therefore incorporated about
300 batch cycles. 

\section{Conclusion}

We have presented here a complete description of the apparatus used in the experimental measurement of the electric dipole moment of the neutron at ILL, Grenoble, and discussed many aspects of the hardware that could have introduced systematic errors into the results.  The equipment was used to take data from 1996 until 2002, at which time it was decommissioned.    At the time of writing, this experiment has provided the world's most sensitive limit on the neutron EDM.

\acknowledgments

We acknowledge with gratitude the contributions of  Prof. N.F.\ Ramsey, original pioneer of the entire
series of neutron EDM experiments, for his invaluable interest and support 
throughout this long program of work. During the many years of research and
development that lie behind this publication,
there have been many significant individual contributions. Here, we would
like to acknowledge the contributions made by Ph.D.\ students from the
University of Sussex to the previous neutron EDM experiment, upon which the apparatus for this measurement was strongly based and in whose theses detailed prescriptions for many of the early experimental
techniques can be found, namely A.R. Taylor (1977); T. Sumner (1979); S.M. Burnett
(1982); P. Franks (1986); P.M.C. de Miranda (1987); N. Crampin (1989); and D.J.
Richardson (1989).  One of us (DJM) benefitted from an ILL studentship.  Much of the detailed on-site work with neutrons at
the ILL was carried out by several generations of Research Fellows and
engineers, and in this regard we would particularly like to thank R. Golub,
J. Morse, H. Prosper and D.T.
Thompson.  Prof.\ L.\ Hunter alerted us to the possibility of a geometric-phase-induced systematic effect caused by the ``conspiracy'' between $\dBzdz$ and the $\vxE$ effect.  We would like to thank Prof.\ C.\ Cohen-Tannoudji for useful discussion relating to frequency shifts caused by the mercury reading light.  Prof. A. Serebrov of PNPI assisted us greatly with the provision of high-quality neutron guides.  Prof. Lobashev, also of PNPI, helped us to obtain our first quartz-walled neutron containment vessel, and Natasha Ivanov helped us to obtain various optical components.  The experiments would not have been possible at all without
the cooperation and provision of neutron facilities by the ILL itself. This
work has been funded by the U.K. Science and Technology Facilities Council (STFC) and by its predecessor, the Particle Physics and Astronomy Research Council (PPARC).  Much of the early magnetometry work, in particular that carried out at the University of Washington, was supported by NSF grant PHY-8711762.

\bibliography{neutron_EDM}

\end{document}